\documentclass[twocolumn,showpacs,preprintnumbers,amssymb,amsmath,superscriptaddress,letterpaper]{revtex4}[10pt]

\usepackage{graphicx}
\usepackage{dcolumn}
\usepackage{bm}
\usepackage{pstricks}
\usepackage{color}
\usepackage{epsfig}

\newcommand{\be}{\begin{equation}}
\newcommand{\ee}{\end{equation}}
\newcommand{\bea}{\begin{eqnarray}}
\newcommand{\eea}{\end{eqnarray}}

\newcommand{\s}{{\rm ~s}}

\newcommand{\cm}{{\rm ~cm}}

\newcommand{\microG}{\mu{\rm G}}

\newcommand{\GeV}{{\rm ~GeV}}

\newcommand{\kpc}{{\rm ~kpc}}

\def\ltap{\ \raise.3ex\hbox{$<$\kern-.75em\lower1ex\hbox{$\sim$}}\ }
\def\gtap{\ \raise.3ex\hbox{$>$\kern-.75em\lower1ex\hbox{$\sim$}}\ }

\newcommand{\brem}{bremsstrahlung }

\newcommand{\epp}{e^\pm}

\begin{document}

\title{A New Approach to Searching for Dark Matter Signals in Fermi-LAT Gamma Rays}

\author{Spencer Chang}
\affiliation{Physics Department, University of California Davis, Davis, CA 95616}

\author{Lisa Goodenough}
\affiliation{Center for Cosmology and Particle Physics, Department of Physics, New York University, 
New York, NY 10003}

\date{\today}

\begin{abstract}

Several cosmic ray experiments have measured excesses in electrons and positrons, relative to standard backgrounds, for energies from $\sim$ 10 GeV - 1 TeV.  These excesses could be due to new astrophysical sources, but an explanation in which the electrons and positrons are dark matter annihilation or decay products is also consistent.  Fortunately, the Fermi-LAT diffuse gamma ray measurements can further test these models, since the electrons and positrons produce gamma rays in their interactions in the interstellar medium.  Although the dark matter gamma ray signal consistent with the local electron and positron measurements should be quite large, as we review, there are substantial uncertainties in the modeling of diffuse backgrounds and, additionally, experimental uncertainties that make it difficult to claim a dark matter discovery.  In this paper, we introduce an alternative method for understanding the diffuse gamma ray spectrum in which we take the intensity ratio in each energy bin of two different regions of the sky, thereby canceling common systematic uncertainties.  For many spectra, this ratio fits well to a power law with a single break in energy.  The two measured exponent indices are a robust discriminant between candidate models, and we demonstrate that dark matter annihilation scenarios can predict index values that require ``extreme'' parameters for background-only explanations.
  
\end{abstract}
\maketitle

\section{Introduction}
\label{sec:intro}

Since the introduction of dark matter into the scientific consciousness by Zwicky over 85 years ago, its nature has remained elusive.  Though the existence of dark matter (DM) has been verified by numerous observations, we are still ignorant about its most basic properties, such as its mass and its interactions with standard model particles.  A number of direct and indirect detection experiments are currently underway in an effort to change this, and several cosmic ray experiments have released interesting results over the past year.  If these results are interpreted as signals of dark matter, then they lead to some surprising conclusions about this basic component of our universe.
 
PAMELA \cite{Adriani:2008zr, PAMELA2, PAMELA} has measured a pronounced increase in the positron fraction above 10 GeV.  The secondary positron spectrum, which originates from cosmic ray (CR) interactions with the interstellar gas, is softer than the primary electron spectrum, so in the absence of additional primary sources, the positron fraction was expected to drop with energy.  The upturn in the positron fraction suggests the existence of a new primary source of positrons with energies in the range $\sim 10-100$ GeV.  
Astrophysical sources such as pulsars \cite{pulsars,pulsars2,2001A&A...368.1063Z,Hooper:2008kg,Yuksel:2008rf,Profumo:2008ms,Malyshev:2009tw,Kawanaka:2009dk,Grasso:2009ma} or effects from standard cosmic ray sources (secondary $\epp$ production in the shocks of supernova remnants (SNRs) \cite{Blasi:2009hv,Blasi:2009bd} or local inhomogeneities in their numbers \cite{Piran:2009tx}) offer explanations for the PAMELA excess, but a dark matter origin is an exciting possibility. 

The Fermi collaboration \cite{Abdo:2009zk} has measured the $e^+\!+e^-$ flux from $\sim\!20$ and $\sim\!900$ GeV.  Between 100 and 500 GeV, the spectrum is well described by a power law with an index that is harder than what is predicted by conventional theories of electron propagation in the Galaxy.  Additionally, the Fermi data suggests a break at the highest energies. HESS \cite{Collaboration:2008aaa,Aharonian:2009ah} has confirmed a break in the spectrum around $\sim \!1$ TeV.  The results of these two experiments are consistent with a new, primary source contributing to the local fluxes of electrons and positrons in the energy range $10-1000$ GeV.  As with the PAMELA results, these may be a sign of dark matter.

Regardless of the details of the production method, a population of high energy electrons and positrons will give rise to gamma rays with energies up to several hundred GeV through the process of inverse Compton scattering (ICS) off the low energy photons in the interstellar radiation field (ISRF).  The Fermi Gamma-ray Space Telescope experiment is currently making all-sky measurements of gamma rays from tens of MeV to more than 300 GeV with precise energy resolution and unprecedented angular resolution.  Fermi is therefore expected to detect the gamma ray products of the surprising new population of $\epp$. The challenge will be to distinguish the gamma ray signal of the newly discovered high energy $\epp$ from the conventional background signal, so that the nature of the source can be ascertained.

\vspace{\baselineskip}

Several theories have been proposed to provide consistent explanations of both the PAMELA and Fermi results in terms of dark matter annihilation or decay \cite{ArkaniHamed:2008qn,Chen:2008yi,Nelson:2008hj,Cholis:2008qq,Nomura:2008ru,Harnik:2008uu,Bai:2008jt,Fox:2008kb,Ponton:2008zv,Chen:2008qs,Ibe:2008ye,Chun:2008by,Arvanitaki:2008hq,Grajek:2008pg,Shirai:2009kh,Mardon:2009gw}.  While the details of the various models differ, explanations of the high energy $\epp$ signals as arising from DM \emph{annihilation} have one thing in common:  they require the annihilation rate to be 2-3 orders of magnitude larger than that of a conventional thermal relic, i.e., the effective thermally averaged annihilation cross section today needs to be $\sim \!100-1000$ times larger than $3\times 10^{-26}\; \rm cm^{3} s^{-1}$, the expected thermally averaged annihilation cross section for a WIMP with mass $m_{\chi}\!\sim\! 500$ GeV.  While this poses theoretical challenges, it is good news on the experimental side, as it makes observing a dark matter signal at other indirect detection experiments more likely.  Specifically, if the additional high energy electrons and positrons observed at PAMELA and Fermi are a DM annihilation signal, then the observed large annihilation rate makes the prospect of measuring the ICS signal promising.

\vspace{\baselineskip}

Many analyses of the DM gamma ray signal have been done \cite{Serpico:2008ga,Martinez:2009jh,Baltz:1999ra,Kuhlen:2008aw,Baltz:2008wd,Bergstrom:2001jj,Elsaesser:2004ap,Kuhlen:2009jv} focusing on searches, for example, in dark matter subhalos, dwarf galaxies, and in the extragalactic diffuse emission.  The galactic diffuse emission has long been the focus of dark matter gamma ray searches \cite{SiegalGaskins:2008ge,Dodelson:2007gd,Pieri:2009je,Springel:2008by,Cholis:2009gv}.
In spite of the large dark matter signal rate associated with the PAMELA and Fermi results, given an observed diffuse gamma ray spectrum, it is not straightforward to show that it requires a dark matter contribution.  Unfortunately, theoretical uncertainties in the background gamma ray spectra, along with the statistical and systematic uncertainties of the experiment, ensure that this is a challenging endeavor.   Given the situation, it is useful to consider other approaches to analyzing the data.  In this paper, we investigate an alternative method that removes some of the common systematic uncertainties that plague interpretations of gamma ray spectra.  The idea is to consider the spectra for two different sky regions and take their intensity ratio in each energy bin, whereby systematic uncertainties that are energy dependent will cancel.  Moreover, fitting this ratio's energy dependence to a power law with a single break gives two measured exponent indices that can be robustly compared with different candidate models.  As we demonstrate, for gamma ray spectra due to some dark matter models which fit the Fermi electron spectrum and early gamma ray data, the power law fit gives values for the indices that are hard to reproduce with background alone, unless rather extreme conditions are satisfied by the background.  Since these indices are well measured and free from systematics, a signal consistent with the dark matter model can be more rigorously argued as evidence for its observation.           

The rest of the paper is organized as follows:  in Section~\ref{sec:gammas}, we give a review of the different processes and components of the diffuse gamma ray spectrum.  In Section~\ref{sec:analysis}, we state our calculation methods, discuss our signal and backgrounds, and then describe in detail the ratio method.  Readers who wish to skip these preliminaries should start at Section~\ref{sec:details}.  In Section~\ref{sec:results}, we present our main results and in Section~\ref{sec:conclusions}, we conclude.

\section{Gamma Ray Spectrum \label{sec:gammas}}

\begin{figure}
\begin{center}
\includegraphics[width=.48\textwidth]{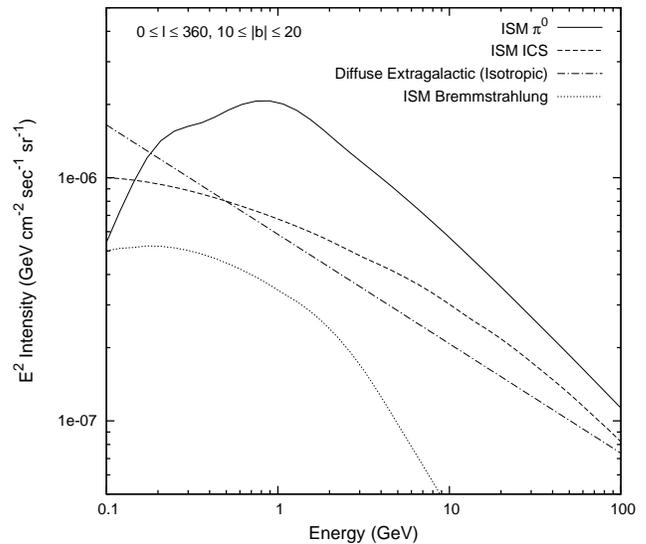}
\end{center}
\caption{The components of the gamma ray sky: $\pi^0$ (solid), ICS (dashed), bremsstrahlung (dotted), extragalactic isotropic (dot-dashed).  The point source and DM contributions are not shown.}
\label{fig:GRcomp}
\end{figure}

The gamma ray sky as measured by Fermi is made up of several components: the $\pi^0$, inverse Compton scattering and \brem components coming from interactions of cosmic rays with the interstellar medium (ISM), as well as the point source and extragalactic isotropic components, as shown in Figure~\ref{fig:GRcomp}.  Additionally, dark matter is expected to contribute, either through direct production of gamma rays as it annihilates or decays, or through the interactions of its annihilation (decay) products with the ISM.

The spectra of the various components can vary with the direction of observation, as the spectral shape and magnitude of a given component may be position dependent.  The contribution from the prompt decay of neutral pions ($\pi^0 \rightarrow \gamma \gamma$) produced in the \emph{p-p} collisions of cosmic ray (CR) protons with the interstellar gas traces the distribution of CR protons and the gas distribution, which is peaked toward $z=0$.

CR electrons, which include both the primaries accelerated by supernova shocks and the secondaries produced in the \emph{p-p} collisions of CR's with the interstellar gas, propagate through the Galaxy losing energy through  through synchrotron radiation in their interactions with the Galactic magnetic field as well as inverse Compton scattering off of the photons of the Galactic interstellar radiation field (ISRF).  Consequently, the distribution of ICS gamma rays is related to the distribution of the CR electron and the ISRF densities.  The interstellar radiation field is composed of three sources: the low energy, isotropic CMB photons, the optical photons due to starlight, which are concentrated near the Galactic Plane (GP), and the infrared (IR) photons emitted by dust that has absorbed optical photons.  The IR and optical components of ICS are thus very strongly peaked toward the GP where the starlight and dust are located.

There is a diffuse isotropic component of gamma rays from unresolved sources such as active galactic nuclei and from diffuse emission processes outside of the Galaxy.  A measurement of this extragalactic emission has been made by the Fermi collaboration and is compatible with a power law index of $-2.41$ between 200 MeV and 100 GeV \cite{Abdo:2010dk} as displayed in Figure~\ref{fig:GRcomp}.  Additionally, there is the contribution from standard gamma ray point sources, such as pulsars and blazars, which are more concentrated near the Galactic Plane.

Models of dark matter with high energy electrons and positrons as annihilation or decay products predict a diffuse contribution to the gamma ray signal from \brem and ICS \cite{Cholis:2008wq,Zhang:2008tb,Borriello:2009fa,Cirelli:2009vg,Regis:2009md,Belikov:2009cx,Meade:2009iu} and a prompt contribution from final state radiation (internal bremsstrahlung) \cite{Beacom:2004pe,Bergstrom:2004cy,Birkedal:2005ep,Mack:2008wu,Bergstrom:2008gr,Bertone:2008xr,Bergstrom:2008ag,Meade:2009rb,Mardon:2009rc,Meade:2009iu}.  The prompt component has the same distribution as the \emph{square} of the dark matter density in models of DM annihilation (or as the dark matter density itself in models of DM decay).  Electrons and positrons lose energy through ICS with an energy loss rate that is proportional to the square of their energy.  Therefore, the characteristic distance they diffuse before losing most of their energy is much smaller for the highest energy electrons than for lower energy electrons.  As a result, the highest energy ICS photons have a distribution that is similar to that of the prompt DM gammas, while the lower energy ICS photons exhibit a more diffuse distribution.

\section{Analysis \label{sec:analysis}}

\subsection{Galprop}

We use the GALPROP code (version v50p) of Moskalenko and Strong to numerically propagate cosmic rays and calculate the resulting gamma ray production in the Galaxy \footnote{The source code and a complete description of the GALPROP program are available on the GALPROP homepage \texttt{http://galprop.stanford.edu}.}.   In the spatial dynamics it includes diffusion resulting from cosmic rays scattering on magnetohydrodynamic waves, while in momentum space, energy losses from ionization, bremsstrahlung, inverse Compton scattering and synchrotron radiation are included \cite{Strong:2007nh,Moskalenko:1999sb}.

Using the most recent measurements of the source abundances for the primary CR species, including nuclei, electrons and gamma rays, GALPROP propagates the primary CR's through the Galaxy, iteratively computes the resulting spallation source functions for all species, and then propagates the full (primary + secondary) source function for each species until a converging result is obtained, assuming free escape of the particles as the spatial boundary condition \cite{Strong:1999sv, Galprop1}.

The calculation of the ICS gamma rays requires knowledge of the interstellar radiation field.  For a full discussion of the ISRF as modeled in GALPROP see \cite{Porter:2005qx}.  Additionally, the atomic (HI), molecular ($\rm H_2$), and ionized (HII) hydrogen gas distributions are needed to compute the gamma rays from bremsstrahlung and $\pi^0$-decay.  The realization of these distributions in GALPROP is explained in detail in \cite{Strong:1998pw}.

For our calculations, we assume cylindrical symmetry, as well as mirror symmetry with respect to the Galactic Plane.  Our standard diffusion zone has a radius of $20$ kpc and a height of 
$\pm4 \kpc$.  We take as our standard diffusion constant $K(E)=5.30\times10^{28} (E/4\GeV)^{0.43} \cm^2 \s^{-1}$.  These choices of diffusion parameters give values of the ratios B/C, sub-Fe/Fe, and $\rm^{10}Be/^{9}Be$ that are in agreement with the local data.

We use a Galactic magnetic field parametrized as follows:
\be
\label{eqn:magfield}
\langle B^2\rangle^{1/2} = B_0 \exp\left[-\frac{(r-R_{\odot})}{r_B} - \frac{|z|}{z_B}\right]
\ee
where $r_B=10\kpc$ is the scale radius, $z_B=2\kpc$ is the scale height, and $B_0 = 5.0\, \microG$ is the local value of the field.
We note that it is inconsistent to assume that the Galactic magnetic field is \textit{inhomogeneous} and \textit{anisotropic}, while simultaneously assuming that the diffusion coefficient is \textit{homogeneous} and \textit{isotropic}.  However, the most current, publicly available version of GALPROP does not have the capability to treat diffusion as spatially dependent.  Moreover, we use the spatially dependent magnetic field of Eq.~\ref{eqn:magfield} in order to best model the spatial dependence of energy losses of electrons and positrons in the Galaxy.

\subsection{DM Parameters}
We take the cusped Einasto profile as defined in Merritt et al \cite{Merritt:2005} as our ``benchmark" dark matter density profile.  Additionally, we consider a cored Isothermal profile as defined in \cite{Moskalenko:1999sb} as an example of a cored DM density profile.  These are defined as follows:

\begin{align} 
\rho(r) = & \;\rho_0 \exp\left[-\frac{2}{\alpha} \left(\frac{r^\alpha - R_\odot^\alpha}{r_{-2}^\alpha} \right) \right] & \mathrm{Einasto} \\ 
\rho(r) = & \;\rho_0 \frac{r_c^2 + R_\odot^2}{r_c^2 + r^2} & \mathrm{Cored\; Isothermal}
\end{align}

\noindent $R_{\odot}=8.5 \kpc$ is the solar distance from the Galactic center, $0.13\leq\alpha\leq0.22$ is the parameter that defines the cuspiness of the Einasto profile (we use 0.17), $r_{-2}=25 \kpc$ is the radius at which the logarithmic slope of the Einasto profile is $-2$, and $r_c=2.8 \kpc$ is the core radius for the cored Isothermal profile.  We take $\rho_0 = 0.3 \GeV \cm^{-3}$  as the local value of the DM mass density.

We define the boost factor ($BF$) of the dark matter signal as the ratio of the thermally averaged annihilation cross section needed to fit a set of data, $\langle\sigma_{ann}\vert v\vert\rangle_{fit}$, to $3\times 10^{-26}\; \rm cm^{3} s^{-1}$, the expected thermally averaged annihilation cross section for a WIMP with mass $m_{\chi}\sim 500\; \rm GeV$,

\be
BF=\frac{\langle\sigma_{ann}\vert v\vert\rangle_{fit}}{3\times 10^{-26}\;  \rm cm^{3} s^{-1}}.
\label{eq:boost_eq}
\ee
Here, the boost factor represents the enhancement in the dark matter annihilation rate coming from an enhanced annihilation cross-section.  The typical annihilation rate needed to explain the PAMELA and Fermi data is 1-2 orders of magnitude larger (depending on the annihilation channel) than what is expected for a thermal relic with a relic density of $\Omega h^2 = 0.105$.  The enhancement in the rate due to the effects of substructure is expected to be $\sim2$, so most of the needed enhancement is expected to come from particle physics, i.e. from the annihilation cross-section.  We assume the annihilation cross-section is spatially isotropic, though this may not be the case.  Indeed, in theories of Sommerfeld-enhanced annihilation \cite{Hisano:2004ds,Cirelli:2007xd, ArkaniHamed:2008qp} in which the cross-section depends on the relative velocity of the particles participating in the annihilation, the cross-section could be spatially dependent, as the velocity dispersion of the dark matter halo varies with distance from the Galactic Center.  However, there is disagreement on the nature of the spatial variation of the velocity dispersion.  For example, some simulations suggest a larger velocity dispersion in the Galactic Center than at the solar position, while some suggest a smaller velocity dispersion.  Therefore, we take the annihilation cross-section, and thus the $BF$, to be spatially independent.

\subsection{Difficulties with Seeing the DM signal at Fermi}

As mentioned earlier, the dark matter gamma ray signal detectable at Fermi consists of several components; the diffuse ICS and bremsstrahlung signals from the interactions of the electrons and positrons produced in annihilations with the interstellar radiation field and the interstellar gas, respectively, and the prompt signals from internal bremsstrahlung off of charged annihilation products, i.e final state radiation (FSR), and from the decay of neutral pions produced in the annihilations \footnote{Additionally, there can be a prompt component due to the annihilation mode $\chi \chi \rightarrow \gamma \gamma$, though the branching ratio to this mode is usually very small due to loop suppression}.

For the dark matter scenarios we consider here, the diffuse bremsstrahlung signal is 1-2 (or more) orders of magnitude smaller than the diffuse ICS signal, so we neglect it in our calculation. The FSR signal becomes relevant only at energies of order the DM mass (above the currently stated Fermi range of 10 MeV-300 GeV), and then only for certain annihilation modes, DM masses, and boost factors.  Therefore, we neglect the contribution of final state radiation to the dark matter gamma ray spectrum.  Furthermore, we do not consider the effects of a prompt $\pi^0$-decay component on the DM signal.  Such a signal has a pronounced bump at high energies that extends well into the Fermi energy range and would be easily detectable for the relatively large branching ratios that would generically occur.  However, the PAMELA collaboration's measurement of the antiproton fraction \cite{Adriani:2008zq} places limits on the branching ratio to hadrons which severely constrain the flux of $\pi^0$'s coming from hadronic annihilation channels.  Fortunately, such suppressed or nonexistent $\pi^0$ production can occur naturally in many dark matter models.  Additionally, the total gamma ray flux as already measured by EGRET \cite{Hunter:1997we,Sreekumar:1997un} and HESS \cite{Aharonian:2004wa,Aharonian:2006wh} places constraints on a prompt component of any type \cite{Cholis:2008wq,Bell:2008vx,Bertone:2008xr,Bergstrom:2008ag,Meade:2009rb,Mardon:2009rc,Meade:2009iu}.  In light of all of this, for the dark matter signal, we henceforth consider the gamma rays produced from inverse Compton scattering of $\epp$ produced in DM annihilations.

\vspace{\baselineskip}

The Fermi $e^+\! + e^-$ data suggest that $M_{DM} \gtap 1000 \GeV$ and that the spectra of $\epp$ from DM annihilation are fairly soft.  For example, direct annihilation into electrons gives too hard a spectrum, but channels with softer electrons, like $\chi \chi \rightarrow \mu^+ \mu^-$ and $\chi \chi \rightarrow \phi \phi$ followed by $\phi \rightarrow l^+ l^-$, give good fits.  Given the ``large" annihilation cross sections these channels need to fit the Fermi and PAMELA data, $BF \gtap 200$, one might expect the DM gamma ray signal to be easily detected at Fermi.  However, the gamma ray signal may not be as large as the ``local" signals suggest.  The boost factor is highly position dependent, as it depends both on the extent of substructure in a particular region of the Galaxy and on the velocity dispersion of the dark matter particles, and the line-of-sight average may be significantly lower than the local value.  

Even with a large DM signal, it may prove difficult to disentangle the DM signal from the flux due to more mundane sources.  The Galactic Center region, where the DM gamma ray flux is peaked due to the $\rho_{DM}^2(\mathbf{r})$ dependence of the signal, has large contamination by gamma rays from unidentified point sources.  A measurement of the DM component in this region depends on an accurate determination of the point source contribution.

Perhaps the largest barrier to the detection of a DM signal is the uncertainty in the diffuse gamma ray background, the $\pi^0$-decay gammas and the ICS and bremsstrahlung gammas from primary CR $e^-$ and secondary CR $\epp$.  The diffuse background has uncertainties that arise from our incomplete understanding of cosmic ray propagation, and from our inability to accurately measure the spectrum of cosmic rays throughout the Galaxy, the distribution of gasses in the interstellar medium, the Galactic magnetic field, and the distribution of radiation energy density, which is closely tied to the distribution of dust in the Galaxy.  Within these uncertainties, in many regions of the gamma ray sky, especially those less prone to source contamination, it is often possible to take a spectrum that looks to contain a significant dark matter component and fit it to background alone.  Thus, it is very difficult to use a single interesting gamma ray spectrum as evidence for dark matter.

\subsection{The Details \label{sec:details}}

We take the background gamma ray flux to be the sum of the contributions from $\pi^0$ decay, ICS, and \brem and henceforth call this the ISM background.  The ICS and \brem are themselves composed of gammas arising from the interactions of the primary CR $e^-$ with the ISM, we call these the ``leptonic" component of the ISM background, and of gammas arising from the interactions of the secondary CR $\epp$ with the ISM.  Since these secondary $\epp$ are produced in the interactions of cosmic ray protons with the ISM, we call the sum of these with the $\pi^0$ component the ``hadronic" component of the ISM background.  The leptonic component of the ISM background depends on the spectrum of primary electrons, while the hadronic component depends on the spectrum of primary protons.

While there are many DM annihilation modes that give good fits to the Fermi and PAMELA data, we use the annihilation mode through a light mediator $\phi$ into muons, $\chi \chi \rightarrow \phi \phi \rightarrow 2\mu^+ 2\mu^-$, as a template dark matter signal.  Modes producing much harder $\epp$ spectra do not fit the Fermi $e^+\!+ e^-$ data as well, and softer modes produce similar $\epp$ spectra in the $\sim 100-800$ GeV range (though for different boost factors), as they must in order to fit the Fermi data.  Since the populations of DM $\epp$ are similar, the dark matter ICS signals are therefore similar in the Fermi energy range.

\vspace{\baselineskip}

We calculate the gamma ray spectrum for several scenarios, some with and some without a dark matter component.  The ISM hadronic and leptonic components are generated by primary proton and electron spectra that are consistent with local data.  We will describe in more detail these local fits below.  We fit the total flux, ISM + DM ICS, to the gamma ray data recently presented by the Fermi collaboration \cite{ACKERMANNTALK}.  The data are given for three regions, ($ 0^\circ < \ell < 360^\circ $, $ 10^\circ < |b| < 20^\circ$), ($ 0^\circ < \ell < 360^\circ $, $ 20^\circ < |b| < 60^\circ $), and ($ 0^\circ < \ell < 360^\circ $, $ 60^\circ < |b|$), which henceforth will be referred to as the Intermediate Latitudes 1 (IL1) region, the Intermediate Latitudes 2 (IL2) region, and the Polar region, respectively, or collectively as the ``regions".  We assume that the extragalactic isotropic component and the point source component are well measured by the Fermi collaboration, so we fit only to the Galactic diffuse flux as defined in the presentation of the Fermi ``regions" data.

We constrain the primary proton spectra to be in agreement with the local proton measurements by AMS-01 \cite{AMS}, BESS \cite{BESS}, and IMAX \cite {IMAX} above 1 GeV.  Accordingly, we take the best-fit power law index below 9 GeV to be $-1.40$.  Above 9 GeV, we allow the power law index to vary such that the resulting spectrum does not exceed the local data by more than $3 \sigma$.  The indices for the fits to data of the hardest and softest allowed spectra are $-2.50$ and $-3.00$, respectively, while a very good fit is achieved for $\sim -2.65$.  The softening of the proton spectrum upon propagation is dominated by energy loss (diffusion) processes at low (high) energies.  With the propagation and energy loss parameters we use in our calculations, the proton spectrum softens upon propagation by an average of $\sim -0.35$ for energies in the range 10-500 GeV.  Therefore, we take the range of acceptable power law indices for the injection spectrum of primary protons to be $-2.15$ to $-2.65$ above 9 GeV.

Similarly, we constrain the spectrum of primary electrons in the range $\sim \! 10-1000$ GeV using the recent Fermi data \cite{Abdo:2009zk}.  Fitting to the data between $\sim 73$ GeV and $\sim 380$ GeV, we find that the range of indices $-2.83$ to $-3.20$ gives spectra that are consistent with the data such that no point exceeds any Fermi data point by more than $3 \sigma$  \footnote{The fits to all Fermi data up to $\sim 380$ GeV constrain the range of indices to be slightly smaller at $-2.94$ to $-3.14$.  However, the data seems best fit by a power law with a break around $\sim 73$ GeV, so we choose to fit the higher energy region.}.  We stress that these constraints arise from fitting only the primary electrons to the $e^+\!+ e^-$ data.  Inclusion of additional components in the flux, such as a dark matter or pulsar component, allows for considerable softening of the allowed primary electron spectrum.  However, we find that our conclusions are unchanged by considering a softer primary electron spectrum, so we take $-3.20$ to be the softest allowed power law index.  Additionally, we note that we have not included the fluxes of secondary electrons and positrons in our fits, which are 1-2 orders of magnitude smaller than the flux of primary $e^-$'s between 10 GeV and 100 GeV.  Including these would result in a small change of the spectral range.  The electron spectrum softens upon propagation by $\sim -0.65$ in the energy range of interest for our choice of propagation and energy loss parameters, so we take the range of acceptable power law indices for the injection spectrum of electrons to be $-2.00$ to $-2.55$.  All power law indices quoted for the electron and proton spectra will be those for the unpropagated spectra.

\begin{figure}
\begin{center}
\includegraphics[width=.48\textwidth]{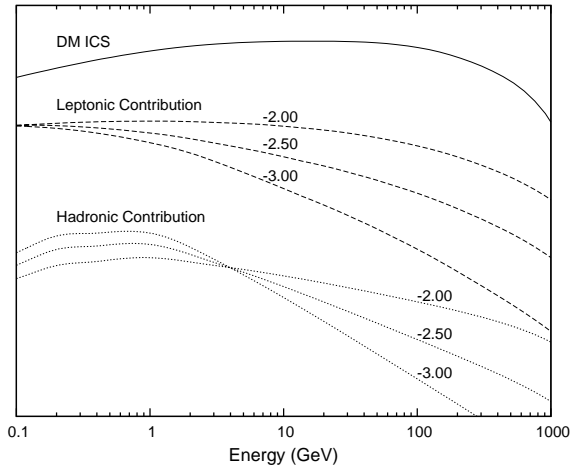}
\end{center}
\caption{Plots for the contributions to the box region ($ 10^\circ <| \ell | < 20^\circ $, $ 10^\circ < |b| < 20^\circ$) from dark matter (solid), leptons (dashed), and hadrons (dotted) with arbitrary normalizations.  Above 10 GeV, from top to bottom the index for the primary electrons and protons is $-2.0$, $-2.5$ and $-3.0$.}
\label{fig:BGindices}
\end{figure}

With these choices, we believe we are being conservative and are taking into account all of the realistic region of the parameter space of power law indices for both primary protons and primary electrons, and perhaps some of the unrealistic region.  While we do not think that the local spectrum of primary electrons is well-described by the power law $E^{-2.8}$ above 100 GeV, we do think that it is possible for the electron spectrum to be described as such in some region of the Galaxy, particularly near a CR source of electrons.  Thus, we consider a wide range of hadronic and leptonic ISM spectra that are theoretically and experimentally motivated.  See, Figure~\ref{fig:BGindices} for example spectra.

As mentioned earlier, there are numerous sources of uncertainty in the diffuse background.  For example, a change in the overall normalization of the magnitude of the Galactic magnetic field would increase the power to synchrotron energy losses, thus reducing the ICS signal everywhere, while a change in the parametrization of the magnetic field would lead to changes in the spatial distribution of the ICS.  Similarly, changes in the optical and IR energy densities would result in a different ICS spectrum.  As a final example, an increase in the HI density in a particular region would lead to a corresponding increase in the $\pi^0$ component of the gamma rays.  We do not attempt to address all of these issues here.  Rather we expect that the large range of spectra we consider for the primary electrons and protons serves as a proxy for many of the uncertainties one expects from other sources.

\vspace{\baselineskip}

\begin{figure}
\begin{center}
\includegraphics[width=.48\textwidth]{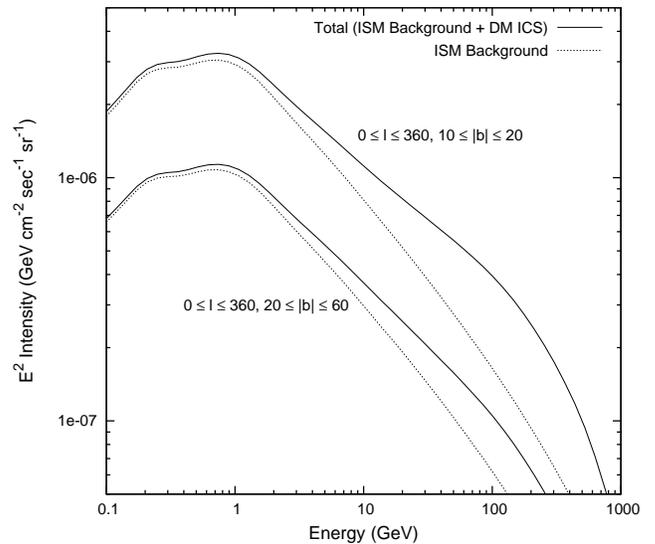}
\end{center}
\caption{The ISM background (dotted) and total, ISM background + DM ICS, (solid) gamma ray signals for the Intermediate Latitudes 1 region ($ 0^\circ < \ell < 360^\circ $, $ 10^\circ < |b| < 20^\circ$) (\emph{upper}) and Intermediate Latitudes 2 region ($0^\circ < \ell < 360^\circ $, $ 20^\circ < |b| < 60^\circ $) (\emph{lower}).  The plots are shown for comparison of the spectral shapes.}
\label{fig:regioncomp}
\end{figure}

\subsection{Taking the Ratio \label{sec:takingtheratio}}
As our discussion on the gamma ray sources has made clear, it is challenging to conclusively determine that a measured gamma ray spectrum {\em requires} a dark matter contribution.  Not only are there uncertainties in the background parameters, but experimental issues, in particular energy normalization and systematic uncertainties, further cloud the issue.  In this regard, it is useful to consider alternative analysis methods that can overcome some of these difficulties.  

The method we explore is to take the ratio in each energy bin of the gamma ray intensity for two different regions of the sky.  This approach has several advantages.  First, as displayed in Figure \ref{fig:GRcomp}, the standard backgrounds typically are dominated by a single contribution with a characteristic shape.  For example, $\pi^0$ gammas are expected to dominate in the 1-10 GeV range in many sky regions.  Thus, for energies where the two regions have the same process generating a majority of the gamma ray flux, the shapes look similar (see Figure \ref{fig:regioncomp}) and give a roughly constant ratio;  this in turn accentuates the energy ranges where the gamma ray contributions are transitioning from one source to another, making new sources like dark matter more prominent.  Second, the systematic uncertainties that are common to a given bin, like the energy dependence of the acceptance, will cancel in the ratio.  On the other hand, the overall energy normalization will still shift this ratio plot to the left or right, but since we  choose to fit the ratio to a power law with a single break, this only shifts the break and not the exponents of the power law.  Similarly, the normalization of the power law is sensitive to overall factors that scale the different regions like any asymmetric exposure or acceptance.  Still, we expect that for a specific choice of the regions for the ratio, a power law fit with a break gives two measurable exponents that are a robust quantitative handle on the source of the gamma rays.

As a specific implementation of this method, we use as our two regions the ``boxes'', $ 10^\circ <| \ell | < 20^\circ $, $ 10^\circ < |b| < 20^\circ$, and the ``strips'', $ |\ell | > 20^\circ $, $ 10^\circ < |b| < 20^\circ$.  These mid-latitude regions have at least two advantages over latitudes closer to the Galactic Plane.  Since they are well separated from the GP, where the density of point sources is highest, a better measurement of the contribution to the gamma ray flux from point sources (resolved and unresolved) can be made within these regions.  Thus, point source contamination is smaller.  For our analysis we assume that point sources have been properly subtracted.  We note that if point source contributions in regions closer to the Galactic Center, for example regions with $5^\circ < |b| < 10^\circ$, can be reliably subtracted, it would be advantageous to use these regions; these regions have a larger dark matter signal, so the effect seen in the ratio would be more pronounced.  

Our choice of mid-latitude regions is also important for distinguishing between a dark matter explanation and astrophysical explanations for the $\epp$ excesses.  The spatial distribution of pulsars has a disk-like morphology with a scale height of 0.2 kpc, which corresponds to a latitude of $\sim 1.3^\circ$.  It has been shown that pulsars give negligible gamma rays above backgrounds at mid-latitudes and thus would not give a significant excess in gamma rays in these regions \cite{Zhang:2008tb,Barger:2009yt}.  Additionally, local inhomogeneities such as nearby pulsars or SNRs would be expected to appear as point or possibly extended gamma rays sources on the sky but are not expected to give a strong signal in each of four box regions relative to the four strip regions.

\begin{figure}
\begin{center}
\includegraphics[width=.48\textwidth]{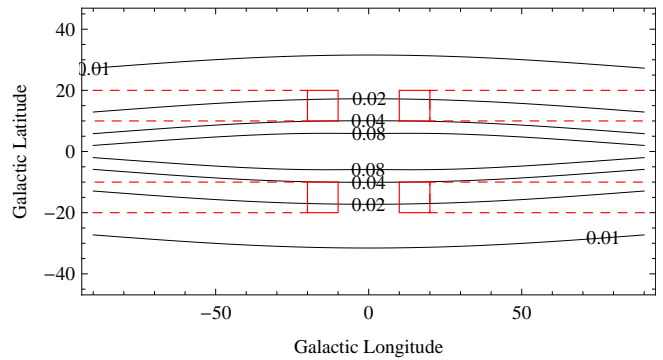}
\end{center}
\caption{A contour plot of the integral along the line-of-sight (los) of the density distribution of supernova remnants as a function of Galactic longitude and latitude.  The integral values have been normalized such that the maximum value is 1.  The box (solid) and strip (dashed) regions used in our analysis are shown. The ratio of the average value of the los integral for the box region to the strip region is 1.39.
}
\label{fig:SNRlosvalues}
\end{figure}

For the case in which the $\epp$ excesses arise from secondary cosmic rays produced at the shock in SNRs \cite{Blasi:2009hv,Blasi:2009bd}, the gamma ray signals  have not been studied, but some comments can be made.  In this scenario there would be a contribution from $\pi^0$ decay (the pions being created in hadronic interactions in the source), which should correlate well with the spatial distribution of CR sources, and a contribution from ICS gammas produced by the hard secondary electrons and positrons.  Standard wisdom holds that the latitude profile of the distribution of SNRs falls off with characteristic distance of a few hundred parsecs. (We assume a source distribution with cylindrical symmetry about $b=0^\circ$ with a scale height of 0.2 kpc in our calculations.)  
In Figure~\ref{fig:SNRlosvalues} we show a contour plot of the line-of-sight (los) integral over the galactic SNR density used in our analysis.  The box and strip regions are outlined for clarity.  Several things are clear from the figure.  First, the los integral over the density of SNRs is fairly constant and the dependence on latitude is stronger than on longitude over our analysis regions.  Second, the average values of the los density integrals are very similar for the boxes and strips; in fact, the ratio of the average value of the los integral for the box region to the strip region is 1.39.  Third, the integral values are very small at mid-latitudes relative to their values within a few degrees of the Galactic Plane, as we would expect for a distribution with a characteristic scale of 0.2 kpc.  These behaviors may serve to differentiate this scenario from a dark matter annihilation scenario.  We note that, as pointed out in \cite{Blasi:2009bd}, the accelerated secondaries explanation of the $\epp$ excesses will also be tested by higher energy data on the antiproton to proton ratio; the ratio is predicted to rise above 100 GeV.  

\section {Results \label{sec:results}}
Our procedure is to take into account the statistical Poisson errors from one year of running at Fermi.  We do not take into account systematic errors, as we argue they should largely cancel in the ratio.  We bin in energies from 100 MeV to 1.887 TeV using logarithmic binning, where the center of each bin is 1.2 times the previous.  The acceptance is taken to be as given in \cite{Atwood:2009ez} multiplied by a relative change in selection (see slide 7 in \cite{ACKERMANNTALK}).  Note that at high energies, the expected observed gamma rays can be less than 1, leading to errors larger than 100\%, which occurs above $\sim$ 300-700 GeV for the box and strip regions.  However, such bins are not crucial to the fits, so we leave them in to illustrate the behavior of the function at high energy.  A more realistic approach would be to look for an optimized binning of these high energy regions.  Finally, we do not take into account any effects due to energy resolution.  

For most of the models, we first fit to the IL1, IL2, and Polar gamma ray data, marginalizing over a common set of indices and normalizations for the proton and electron primaries in these regions.  However, for the dark matter models, we fix the electron primary index to be $-2.45$ since this is used to determine the boost factor necessary for fitting the Fermi electron spectrum, but still allow its normalization to vary.  Next, with the fitted parameters we predict the ratio of the intensity in the boxes over that in the strips and fit to a power law with a single break.  
Since the boxes have more dark matter signal than the strips, the dark matter contribution causes the ratio to increase at high energies.   
We include in the ratios the diffuse isotropic component of the gamma rays as given in \cite{ACKERMANNTALK}, although it is usually unimportant.

\begin{figure}
\begin{center}
\includegraphics[width=.49\textwidth]{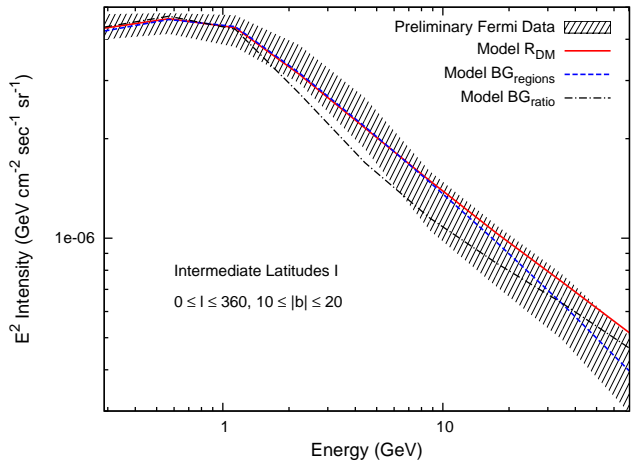}
\end{center}
\caption{The preliminary Fermi-LAT gamma ray spectra in the Intermediate Latitudes 1 region (hatched band) as given in \cite{ACKERMANNTALK}.  Three model fits are plotted:  1) the reference dark matter model $\rm R_{DM}$ (solid), 2) the best background only fit to the three regions $\rm BG_{regions}$ (dashed), and 3) a hard electron spectrum to mimic the dark matter ratio behavior $\rm BG_{ratio}$ (dot-dashed).}
\label{fig:IL1models}
\end{figure}

\begin{figure}
\begin{center}
\includegraphics[width=.48\textwidth]{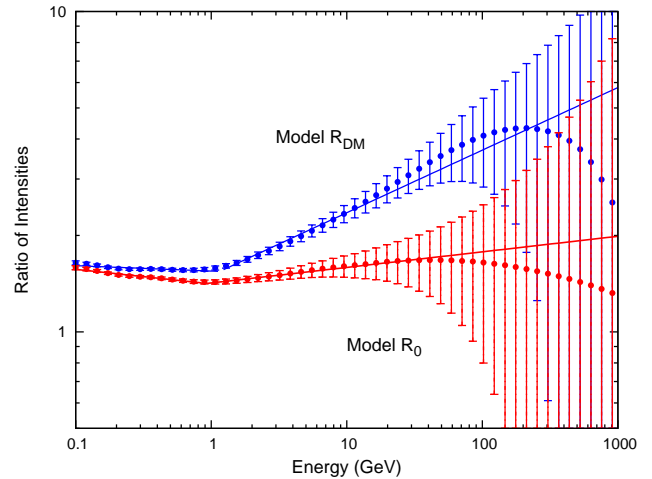}
\end{center}
\caption{Ratio plot for Model $\rm R_{DM}$ (top), 3 TeV dark matter with $BF = 990$, and Model $\rm R_{0}$ (bottom), same background with no dark matter contribution, as a function of energy in GeV.  The fit to a power law with a single break is shown in solid.}
\label{fig:DMvsnoDM}
\end{figure}

Our reference dark matter model, Model $\rm R_{DM}$, is a 3 TeV dark matter particle $\chi$ annihilating in the channel  $\chi \chi \rightarrow \phi \phi \rightarrow 2\mu^+ 2\mu^-$ with a boost factor of 990.  This model gives a very good fit to the Fermi electron spectra, and the spectra in the IL1, IL2 and Polar regions can be fit well by varying the proton's normalization and index in addition to the primary electron normalization.  In particular, this model's spectra in the IL1 region is compared to the best background fit to the regions in Figure~\ref{fig:IL1models}.  In Figure~\ref{fig:DMvsnoDM}, we plot the predicted ratio for this model with and without (Model $\rm R_{0}$) the dark matter contribution.  As can be seen from the plot, both ratios are compatible with a power law with a single break as shown by the solid lines.  For both models, the ratio is roughly constant up to 1 GeV.  On the other hand, from 1-200 GeV  the dark matter ratio is clearly increasing, while for the model without dark matter the ratio stays relatively constant.  In Table~\ref{table:data}, we list the results of the fits to the gamma ray spectra in the IL1, IL2, and Polar regions and the power law fit to the ratio.  For the indices of the power law, we've included the $1\sigma$ error bars by marginalizing the fit over the other parameters of the power law.  The high energy index for the reference DM model is $0.195^{+.010}_{-.005}$, which is well separated from the value for the model with no dark matter, $0.049^{+.011}_{-.014}$.

\begin{figure}
\begin{center}
\includegraphics[width=.48\textwidth]{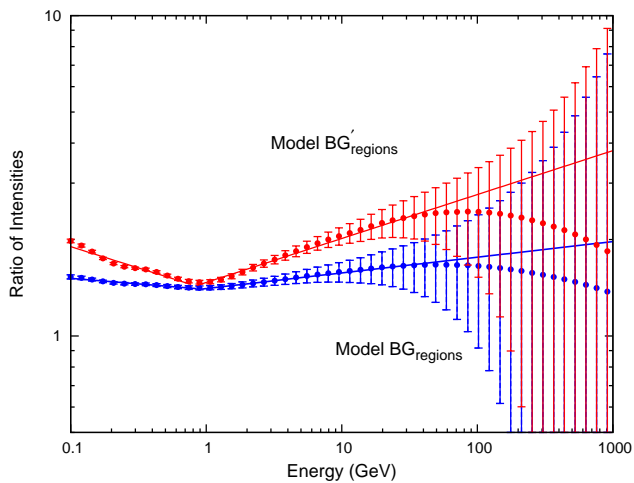}
\end{center}
\caption{Ratio plot for Model $\rm BG_{regions}$ (bottom), the best background fit to the IL1, IL2, and Polar regions, and Model $\rm BG_{regions}'$ (top), same as Model $\rm BG_{regions}$ with a change in relative normalization of 1/2 and 2 for the primary proton and electron spectra in the box region. The fit to a power law with a single break is shown in solid.}
\label{fig:bestBGhpderatio}
\end{figure}

As another comparison, we consider the best fit to the regions with background alone, Model $\rm BG_{regions}$, which has primary proton and electron indices of $-2.38$ and $-2.33$, respectively.  The ratio is shown in the bottom plot of Figure~\ref{fig:bestBGhpderatio}.  The values for the low energy and high energy indices turn out to be similar to Model $\rm R_{0}$, the reference model with dark matter turned off, as listed in Table~\ref{table:data}.  Incidentally, it is also worth mentioning that a background model that is specially designed to fit very well the Fermi electron spectra gives a high energy index of 0.038.  Thus, it seems generic that the best-fit backgrounds to either the Fermi electrons or the Fermi IL1, IL2, and Polar gamma rays do not generate a high energy index that is as large as that for dark matter.  As a side comment, we remind the reader that in Section~\ref{sec:takingtheratio} we briefly discussed the gamma ray signals of the alternative scenario of \cite{Blasi:2009hv} in which the $\epp$ excesses are explained in terms of secondaries accelerated at the shocks in SNRs.  For this scenario, the ratio of the average values of the integral along the line-of-sight of the density distribution of SNRs for the boxes to the strips is 1.39 (see Figure~\ref{fig:SNRlosvalues}).  This value is very similar to the value of the ratio for Model $\rm BG_{regions}$, and we should not be surprised by this - the \emph{sources} are the same in both scenarios.

\setlength{\extrarowheight}{0.2cm}
\begin{table*}[t]
\begin{center}
\begin{tabular}{|c|c|c|c||c|c|c|c|c|}
\hline
& & Proton & Electron & & & Low Energy & High Energy & $E_{break}$ \\[-.4cm]
Model & Description & & & $\frac{\chi^2_{regions}}{\#bins}$ & $\frac{\chi^2_{power}}{\#bins}$ &  &  &  \\[-.4cm] 
& & Index & Index & & & Index & Index & (GeV)\\ \hline
& Reference model & & & & & & & \\[-.5cm]
$\rm R_{DM}$ &  & $-2.47$ & $-2.45$ & 0.23 & 0.33 & $-0.017^{+.004}_{-.003}$ & $0.195^{+.010}_{-.005}$ & 1.15 \\[-.3cm] 
& with dark matter & & & & & & & \\[0cm] \hline
& Reference model & & & & & & & \\[-.5cm]
$\rm R_{0}$ &  & $-2.47$ & $-2.45$ & 0.65 & 0.09 & $-0.046^{+.005}_{-.007}$ & $0.049^{+.011}_{-.014}$ & 0.93\\[-.3cm] 
& without dark matter & & & & & & & \\[0cm] \hline
& Best BG fit to IL1, IL2,  & & & & & & & \\[-.5cm]
$\rm BG_{regions}$ &  & $-2.38$ & $-2.33$ & 0.11 & 0.07 & $-0.034^{+.002}_{-.004}$ & $0.049^{+.010}_{-.012}$ & 0.93 \\[-.3cm]
& and Polar regions & & & & & & & \\[0cm] \hline
& Best BG fit to regions with & & & & & & & \\[-.5cm]
$\rm BG_{regions}'$ &   & $-2.38$ & $-2.33$ & 0.11 & 0.98 & $-0.133^{+.004}_{-.005}$ & $0.137^{+.008}_{-.008}$ & 0.83 \\[-.3cm]
& modified norms in boxes & & & & & & & \\[0cm] \hline
& Hard electron spectrum & & & & & & & \\[-.5cm]
$\rm BG_{ratio}$ &   & $-3.00$ & $-2.00$ & 0.54 & 0.67 & $ -0.000^{+.005}_{-.009} $ & $ 0.204^{+.008}_{-.013} $ & 0.91 \\[-.3cm] 
& fit to ratio & & & & & & & \\[0cm] \hline
& Slightly softer electron& & & & & & & \\[-.5cm]
$\rm BG_{ratio}'$ &   & $-2.52$ & $-2.20$ & 0.13 & 0.91 & $-0.089^{+.004}_{-.005} $ & $ 0.179^{+.008}_{-.008} $ & 0.83 \\[-.3cm]
& spectrum fit to ratio & & & & & & & \\[0cm] \hline
\end{tabular}
\caption{The models with their indices for the primary proton and electron spectra (unpropagated), along with their fit parameters to the IL1, IL2, and Polar regions, and for the ratio plot, the parameters for a fit to a power law with a single break.  \label{table:data}}
\end{center}
\end{table*}

We investigate some variations in the dark matter scenario to determine the effects on the values of the power law fit.  First, we reduce the boost of the dark matter to $(1/2,1/3,1/4)$ which changes the high energy index values to $(0.130,0.107,0.094)$ with only a small change to the low energy index.  Second, we change the mass of the dark matter particle to 1 TeV and 5 TeV, taking, respectively, boosts of 200 and 2000 to give a reasonable fit to the Fermi electron spectrum.  In these cases, we find high energy index values of 0.113 and 0.240, with low indices of $-0.035$ and $-0.014$.  Third, we consider a cored Isothermal dark matter profile with $BF=530$, and we find a high energy index of 0.154.  This is not surprising; the cored Isothermal profile is less cuspy than our standard Einasto profile, and so produces a smaller gamma ray flux near the Galactic Center.  Finally, we alter our diffusion parameters from our standard values.  We consider a diffusion constant of $K(E)=1.29\times10^{28} (E/4\GeV)^{0.40} \cm^2 \s^{-1}$ with a diffusion zone of $|z|\leq 1$, and also a diffusion constant of $K(E)=7.00\times10^{28} (E/4\GeV)^{0.43} \cm^2 \s^{-1}$ with a diffusion zone of $|z| \leq 6$ \footnote{These diffusion parameters give local values of cosmic ray fluxes that are in agreement with local data as described in \cite{Simet:2009ne}.}.  Here we find the high energy index to be $0.035$ and $0.199$, respectively.  This makes physical sense, because for the latitudes we are looking at, $|z| > 1.5$ kpc at $r=0$, hence the smaller diffusion zone is suppressing the electrons from dark matter annihilations that would contribute to the gamma ray flux.  As all of these variations demonstrate, there is some uncertainty in the high energy index value.  However, to be optimistic we continue to use the reference scenario as the target signal we are searching for.     

It is important to determine if it is possible to vary the background to consistently fit the gamma ray spectra in the regions while getting similar index values for the power law fit.  This is one of the benefits of the ratio method, as the power law indices give a quantitative set of numbers to compare to that are not limited by systematics.  For our simplified survey into the possible background variations, we not only vary the primary proton and electron spectra, but also allow for a change in their normalization in the boxes relative to the normalizations used in the IL1, IL2, Polar, and strip regions.  Such normalization changes could be mocked up by variations in Galactic quantities in the boxes relative to the strips.  In fact, to first order, changing the gas density and the strength of the IR component of the ISRF would, respectively, mimic a change in the hadronic and leptonic normalizations.  Based on matching the reference model's high energy index alone, we find that these relative normalizations must be allowed to be as large as a factor of 2.

Our study of these variations shows that there is a tension in the background achieving a good fit to both the regions and the power law fit, while also obtaining both a large high energy index and small low energy index.  The reason for this is due to the particular energy dependencies of the different contributions as shown in Figure~\ref{fig:BGindices}.
Notice that the dark matter ICS contribution (solid) is growing as a function of energy with a broad plateau from a few GeV to 100 GeV, whereas the backgrounds have a much different behavior.  The hadronic contribution (dotted) has a ``camel hump'' plateau from about 200 MeV to 1 GeV with a high energy tail that falls faster for softer values of the primary proton index.  For the leptons (dashed), the contribution is a smoothly falling function.  In general, the region fits favor the hadronic component to dominate at its plateau, with some leptonic contribution at higher energies.  Since the leptonic component does not grow as a function of energy like the dark matter component, to get a large high energy index for the power law fit to the ratio, one needs the lepton component in the boxes to be enhanced relative to the hadronic component.  Due to the shape of the leptonic contribution, it will then also give a sizeable contribution to the low energy region below the peak of the hadronic component.  Thus, the box region's spectra in increasing energy will go from being dominated by leptons, to being a mixture of leptons and hadrons around the hadronic plateau, to again being dominated by leptons.  Since there are three regions of interest, the fit to a power law with a single break will have a poorer $\chi^2$ than the case of dark matter which is dominated by hadrons at low energy and dark matter at high energy.  In the cases where the leptons strongly dominate at low energies, the value of the ratio will actually fall as a function of energy, leading to a noticeable negative low energy index.

\begin{figure}
\begin{center}
\includegraphics[width=.48\textwidth]{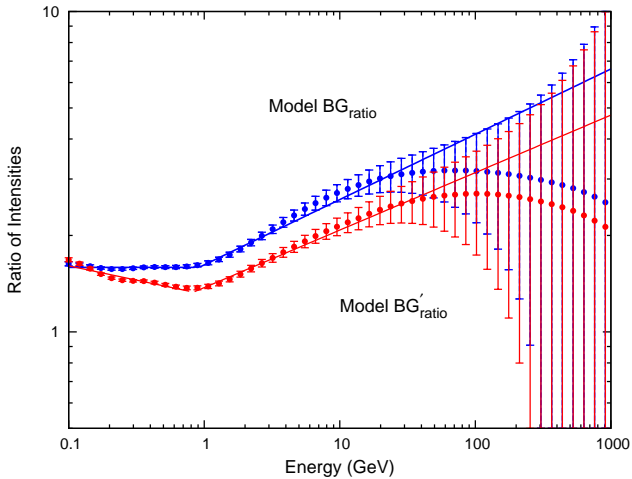}
\end{center}
\caption{Ratio plots for Model $\rm BG_{ratio}$ (top), a background with indices of $-3.00$ and $-2.00$ for primary protons and electrons, respectively, with an increase in relative normalization of 1.6 for the primary electrons in the box region, and for Model $\rm BG_{ratio}'$ (bottom), a background with indices of $-2.52$ and $-2.20$ for primary protons and electrons, with a change in relative normalization of 0.6 and 1.7 in the box region. The fit to a power law with a single break is shown in solid.}
\label{fig:modelAB}
\end{figure}

To illustrate these conclusions, we look at some specific cases.  We consider a case, Model $\rm BG_{regions}'$, in which we take the best fit background to the regions and set the hadronic normalization in the boxes to be 1/2 and the leptonic normalization in the boxes to be 2 times that of the strip region.  The ratio is shown in the top plot of Figure~\ref{fig:bestBGhpderatio}.  It is noticeably falling at low energies and rising at high energies.  As seen in Table~\ref{table:data}, both indices are well separated from reference model $\rm R_{DM}$, and the $\chi^2$ for the power law fit is markedly worse.  However, the low and high energy index can be brought into agreement with the reference model by a judicious choice of indices for the protons and electrons.  Taking softer protons and harder electrons brings out the high energy lepton contribution without as large of a correction at low energies.

In Model $\rm BG_{ratio}$ we take the proton and electron primary indices to be $-3.00$ and $-2.00$, respectively, corresponding to a very hard electron spectrum.  This choice allows for a good balance between decent fits to the gamma ray data in the IL1, IL2, and Polar regions and a ratio that mimics the one for the DM reference model.  For a lepton normalization in the box 1.6 times that in the strip, we get values for the low and high indices similar to the DM reference model, as seen in the top plot of Figure~\ref{fig:modelAB}.  A distinguishing feature of this model is that in the region plots, there is a dip in the spectra around 3 GeV, visible in Figure~\ref{fig:IL1models}, near the transition between hadronic and leptonic contributions.  In addition, the high energy part of the ratio plot has a distinct long plateau starting at 30 GeV, which is considerably different from the reference dark matter model which has a narrow plateau starting at about 200 GeV.  As reflected by the $\chi^2$ fits, with further statistics it will be possible to test if either of these features are truly there.

Finally, we note that the challenge of mimicking the dark matter signal increases if we consider a stricter constraint on the electron index from the fit to the Fermi electron data.  As our Model $\rm BG_{ratio}'$, we take the electron index to be $-2.20$ and the proton index to be $-2.52$ to better fit the region plots.  With a modification of the hadron and lepton normalizations of 0.6 and 1.7 in the box, as shown in the bottom line of Figure~\ref{fig:modelAB}, we get a reasonably large value of the high energy index, but also a large negative low energy index.  

Although this exploration of the background variations has been quite simplistic, it illustrates that the power law fit to the ratio gives useful quantities for models to compare to.  In our study, we have shown that it is necessary to take somewhat extreme parameters for the primary proton and electron spectra to get a ratio that is similar to that for a scenario with a dark matter contribution.   For one, the electron (proton) index typically needs to be on the hard (soft) end of the range of its allowed values.  In addition, their normalizations in the box region have to be as much as a factor of 2 different from that inferred in the strip region.  Although more systematic background studies should be performed, we believe that our general conclusions would still be valid.  This is because for the energies we are interested in, the hadronic and leptonic gamma ray contributions are dominated by a single process.  Reasonable variations of these background components will then largely be captured by our normalization and index changes.  Therefore, we conclude that there is, quite generically, tension in achieving a background model that fits well the gamma ray data for the IL1, IL2, and Polar regions and that has a large high energy index and small low energy index.  Thus, if Fermi were to observe the ratio with a similar behavior as the dark matter reference model, this could be confidently argued as evidence for a dark matter signal in the gamma rays. 

\section{Conclusions \label{sec:conclusions}}

The future Fermi-LAT measurements of the diffuse gamma ray spectrum have the potential to resolve the degeneracies between standard and dark matter explanations of the cosmic electron and positron excesses.  However, background and experimental uncertainties make it difficult to claim a dark matter discovery even if the spectrum is consistent with such a model.  In this regard, it is useful to have many different approaches to this analysis.  In this paper, we suggest a new technique in which the intensity ratio is taken for gamma ray spectra in two different sky regions.  Some systematic uncertainties cancel in the ratio, and by fitting the ratio to a power law with a single break, we find that the two measured power law indices give a robust quantitative handle of the data.  

Operationally, we fit our models to the preliminary Fermi gamma ray spectra in the Intermediate Latitudes 1, Intermediate Latitudes 2, and Polar regions.  We then predict the gamma ray spectra in the box and strip regions that comprise part of the Intermediate Latitudes 1 region.  Computing the ratio of these two spectra, we show that the dark matter scenario has a relatively flat ratio at low energies and an increasing ratio at high energies.  The flatness of the ratio at low energies reflects the similarity of the hadronic contributions to the spectra in the two regions, while the increase in the ratio at high energies reflects the increase in the dark matter contribution at high energies in the box region, which is closer to the center of the Galaxy.  Thus, for the power law fit, the signal has a small low energy index and a large high energy index.  By comparison, background explanations of the Fermi electron and gamma ray data predict a much smaller high energy index.  The background ICS signal, having a spatial dependence that is quite different from that of the DM ICS signal, has a much smaller enhancement in the box region relative to the strip region. 

We believe this method allows for simple comparisons among the various explanations of the positron and electron spectra.  For example, pulsars are known to produce very little gamma ray signal and so would give a flat ratio.  The gamma ray spectra have not been calculated for the supernova remnant explanations \cite{Blasi:2009hv,Blasi:2009bd,Piran:2009tx}, but these scenarios would be amenable to analysis by this method.  However, given the expected SNR density profile, the regions we considered would be less sensitive to these scenarios.  We also investigated the modifications to the background parameters required to mimic the dark matter behavior in gamma rays, and found that quite extreme values of the background parameters are necessary.  First, the electron and proton primary spectra must be, respectively, near the hardest and softest spectra allowed by local data.  Additionally, the electron normalization in the box region must be enhanced relative to the proton normalization by a factor of $\sim 2-4$.  In general, it is difficult for background-only models that fit well the IL1, IL2, and Polar regions' gamma ray data to have a ratio similar to the ratio obtained in the dark matter scenario.  To obtain a high energy index for a background-only model, one must invoke a very hard leptonic component, which is marginally consistent with the local data and possibly inconsistent with expectations for source production.  Given this ability to quantitatively compare models and robustly demonstrate the background's difficulty in imitating the dark matter behavior, a Fermi observation consistent with the dark matter signal can be more robustly argued as evidence for dark matter.

Further studies of this method should be performed to see if these preliminary results are generic.  For example, additional variations (beyond our adjustments of proton and electron primary indices and normalizations) of the background parameters can be studied.  Also, gamma ray predictions for the supernova effects of \cite{Blasi:2009hv,Blasi:2009bd,Piran:2009tx} could be analyzed in this way.  For signals, dark matter models beyond our ``reference'' model can be analyzed, in particular additional annihilation channels, as well as dark matter decay scenarios.  It is possible that the ratio analysis will help make models with much smaller gamma ray signals detectable.  Additionally, the method can also be further optimized.  In this paper, we focused on the intermediate latitude box and strip regions to compute the ratio; other choices may be preferred, while using several regions and taking many ratios could provide even more useful information.  For example, if point source contributions can be reliably subtracted for spectra at lower latitudes, the ratio of these regions would have an even more pronounced dark matter signal and require more extreme backgrounds to explain them.  On the other hand, ratios of two background-dominated regions may be useful in understanding the backgrounds, helping to reduce the background uncertainties.  With future studies and applications along these lines, this new analysis technique may prove useful in explaining cosmic ray excesses and even in providing further evidence for dark matter.

\vskip 0.2 in
\noindent {\bf Acknowledgments}
\vskip 0.05in
\noindent The authors thank Ilias Cholis, Greg Dobler, Doug Finkbeiner, Igor Moskalenko, Simona Murgia, and Neal Weiner for invaluable conversations.  SC is supported in part by the US Department of Energy under contract No. DE-FG02-91ER40674.  LG is supported by the James Arthur Graduate Fellowship in Cosmology, Particle Physics, and Astrophysics at New York University.  Additionally, LG acknowledges the Fermilab Theoretical Astrophysics Group, Batavia, IL 60510 for their hospitality and support during the completion of this work.



\bibliography{ics}

\begin{thebibliography}{87}
\expandafter\ifx\csname natexlab\endcsname\relax\def\natexlab#1{#1}\fi
\expandafter\ifx\csname bibnamefont\endcsname\relax
  \def\bibnamefont#1{#1}\fi
\expandafter\ifx\csname bibfnamefont\endcsname\relax
  \def\bibfnamefont#1{#1}\fi
\expandafter\ifx\csname citenamefont\endcsname\relax
  \def\citenamefont#1{#1}\fi
\expandafter\ifx\csname url\endcsname\relax
  \def\url#1{\texttt{#1}}\fi
\expandafter\ifx\csname urlprefix\endcsname\relax\def\urlprefix{URL }\fi
\providecommand{\bibinfo}[2]{#2}
\providecommand{\eprint}[2][]{\url{#2}}

\bibitem[{\citenamefont{Adriani et~al.}(2008{\natexlab{a}})}]{Adriani:2008zr}
\bibinfo{author}{\bibfnamefont{O.}~\bibnamefont{Adriani}} \bibnamefont{et~al.}
  (\bibinfo{year}{2008}{\natexlab{a}}), \eprint{0810.4995}.

\bibitem[{\citenamefont{Carlson}(2005)}]{PAMELA2}
\bibinfo{author}{\bibfnamefont{P.}~\bibnamefont{Carlson}}
  (\bibinfo{collaboration}{PAMELA}), \bibinfo{journal}{Int. J. Mod. Phys.}
  \textbf{\bibinfo{volume}{A20}}, \bibinfo{pages}{6731} (\bibinfo{year}{2005}).

\bibitem[{\citenamefont{Orsi}(2007)}]{PAMELA}
\bibinfo{author}{\bibfnamefont{S.}~\bibnamefont{Orsi}}
  (\bibinfo{collaboration}{PAMELA}), \bibinfo{journal}{Nucl. Instrum. Meth.}
  \textbf{\bibinfo{volume}{A580}}, \bibinfo{pages}{880} (\bibinfo{year}{2007}).

\bibitem[{\citenamefont{Aharonian et~al.}(1995)\citenamefont{Aharonian, Atoyan,
  and Volk}}]{pulsars}
\bibinfo{author}{\bibfnamefont{F.~A.} \bibnamefont{Aharonian}},
  \bibinfo{author}{\bibfnamefont{A.~M.} \bibnamefont{Atoyan}},
  \bibnamefont{and} \bibinfo{author}{\bibfnamefont{H.~J.} \bibnamefont{Volk}},
  \bibinfo{journal}{Astron. Astrophys.} \textbf{\bibinfo{volume}{294}},
  \bibinfo{pages}{L41} (\bibinfo{year}{1995}).

\bibitem[{\citenamefont{Buesching et~al.}(2008)\citenamefont{Buesching,
  de~Jager, Potgieter, and Venter}}]{pulsars2}
\bibinfo{author}{\bibfnamefont{I.}~\bibnamefont{Buesching}},
  \bibinfo{author}{\bibfnamefont{O.~C.} \bibnamefont{de~Jager}},
  \bibinfo{author}{\bibfnamefont{M.~S.} \bibnamefont{Potgieter}},
  \bibnamefont{and} \bibinfo{author}{\bibfnamefont{C.}~\bibnamefont{Venter}}
  (\bibinfo{year}{2008}), \eprint{arXiv:0804.0220 [astro-ph]}.

\bibitem[{\citenamefont{{Zhang} and {Cheng}}(2001)}]{2001A&A...368.1063Z}
\bibinfo{author}{\bibfnamefont{L.}~\bibnamefont{{Zhang}}} \bibnamefont{and}
  \bibinfo{author}{\bibfnamefont{K.~S.} \bibnamefont{{Cheng}}},
  \bibinfo{journal}{\aap} \textbf{\bibinfo{volume}{368}}, \bibinfo{pages}{1063}
  (\bibinfo{year}{2001}).

\bibitem[{\citenamefont{Hooper et~al.}(2008)\citenamefont{Hooper, Blasi, and
  Serpico}}]{Hooper:2008kg}
\bibinfo{author}{\bibfnamefont{D.}~\bibnamefont{Hooper}},
  \bibinfo{author}{\bibfnamefont{P.}~\bibnamefont{Blasi}}, \bibnamefont{and}
  \bibinfo{author}{\bibfnamefont{P.~D.} \bibnamefont{Serpico}}
  (\bibinfo{year}{2008}), \eprint{0810.1527}.

\bibitem[{\citenamefont{Yuksel et~al.}(2008)\citenamefont{Yuksel, Kistler, and
  Stanev}}]{Yuksel:2008rf}
\bibinfo{author}{\bibfnamefont{H.}~\bibnamefont{Yuksel}},
  \bibinfo{author}{\bibfnamefont{M.~D.} \bibnamefont{Kistler}},
  \bibnamefont{and} \bibinfo{author}{\bibfnamefont{T.}~\bibnamefont{Stanev}}
  (\bibinfo{year}{2008}), \eprint{0810.2784}.

\bibitem[{\citenamefont{Profumo}(2008)}]{Profumo:2008ms}
\bibinfo{author}{\bibfnamefont{S.}~\bibnamefont{Profumo}}
  (\bibinfo{year}{2008}), \eprint{0812.4457}.

\bibitem[{\citenamefont{Malyshev et~al.}(2009)\citenamefont{Malyshev, Cholis,
  and Gelfand}}]{Malyshev:2009tw}
\bibinfo{author}{\bibfnamefont{D.}~\bibnamefont{Malyshev}},
  \bibinfo{author}{\bibfnamefont{I.}~\bibnamefont{Cholis}}, \bibnamefont{and}
  \bibinfo{author}{\bibfnamefont{J.}~\bibnamefont{Gelfand}}
  (\bibinfo{year}{2009}), \eprint{0903.1310}.

\bibitem[{\citenamefont{Kawanaka et~al.}(2009)\citenamefont{Kawanaka, Ioka, and
  Nojiri}}]{Kawanaka:2009dk}
\bibinfo{author}{\bibfnamefont{N.}~\bibnamefont{Kawanaka}},
  \bibinfo{author}{\bibfnamefont{K.}~\bibnamefont{Ioka}}, \bibnamefont{and}
  \bibinfo{author}{\bibfnamefont{M.~M.} \bibnamefont{Nojiri}}
  (\bibinfo{year}{2009}), \eprint{0903.3782}.

\bibitem[{\citenamefont{Grasso et~al.}(2009)}]{Grasso:2009ma}
\bibinfo{author}{\bibfnamefont{D.}~\bibnamefont{Grasso}} \bibnamefont{et~al.}
  (\bibinfo{collaboration}{FERMI-LAT}) (\bibinfo{year}{2009}),
  \eprint{0905.0636}.

\bibitem[{\citenamefont{Blasi}(2009)}]{Blasi:2009hv}
\bibinfo{author}{\bibfnamefont{P.}~\bibnamefont{Blasi}},
  \bibinfo{journal}{Phys. Rev. Lett.} \textbf{\bibinfo{volume}{103}},
  \bibinfo{pages}{051104} (\bibinfo{year}{2009}), \eprint{0903.2794}.

\bibitem[{\citenamefont{Blasi and Serpico}(2009)}]{Blasi:2009bd}
\bibinfo{author}{\bibfnamefont{P.}~\bibnamefont{Blasi}} \bibnamefont{and}
  \bibinfo{author}{\bibfnamefont{P.~D.} \bibnamefont{Serpico}},
  \bibinfo{journal}{Phys. Rev. Lett.} \textbf{\bibinfo{volume}{103}},
  \bibinfo{pages}{081103} (\bibinfo{year}{2009}), \eprint{0904.0871}.

\bibitem[{\citenamefont{Piran et~al.}(2009)\citenamefont{Piran, Shaviv, and
  Nakar}}]{Piran:2009tx}
\bibinfo{author}{\bibfnamefont{T.}~\bibnamefont{Piran}},
  \bibinfo{author}{\bibfnamefont{N.~J.} \bibnamefont{Shaviv}},
  \bibnamefont{and} \bibinfo{author}{\bibfnamefont{E.}~\bibnamefont{Nakar}}
  (\bibinfo{year}{2009}), \eprint{0905.0904}.

\bibitem[{\citenamefont{Abdo et~al.}(2009)}]{Abdo:2009zk}
\bibinfo{author}{\bibfnamefont{A.~A.} \bibnamefont{Abdo}} \bibnamefont{et~al.}
  (\bibinfo{collaboration}{The Fermi LAT}) (\bibinfo{year}{2009}),
  \eprint{0905.0025}.

\bibitem[{\citenamefont{Aharonian et~al.}(2008)}]{Collaboration:2008aaa}
\bibinfo{author}{\bibfnamefont{F.}~\bibnamefont{Aharonian}}
  \bibnamefont{et~al.} (\bibinfo{collaboration}{H.E.S.S.}),
  \bibinfo{journal}{Phys. Rev. Lett.} \textbf{\bibinfo{volume}{101}},
  \bibinfo{pages}{261104} (\bibinfo{year}{2008}), \eprint{0811.3894}.

\bibitem[{\citenamefont{Aharonian}(2009)}]{Aharonian:2009ah}
\bibinfo{author}{\bibfnamefont{H.~E. S. S. C.~F.} \bibnamefont{Aharonian}}
  (\bibinfo{year}{2009}), \eprint{0905.0105}.

\bibitem[{\citenamefont{Arkani-Hamed et~al.}(2009)\citenamefont{Arkani-Hamed,
  Finkbeiner, Slatyer, and Weiner}}]{ArkaniHamed:2008qn}
\bibinfo{author}{\bibfnamefont{N.}~\bibnamefont{Arkani-Hamed}},
  \bibinfo{author}{\bibfnamefont{D.~P.} \bibnamefont{Finkbeiner}},
  \bibinfo{author}{\bibfnamefont{T.~R.} \bibnamefont{Slatyer}},
  \bibnamefont{and} \bibinfo{author}{\bibfnamefont{N.}~\bibnamefont{Weiner}},
  \bibinfo{journal}{Phys. Rev.} \textbf{\bibinfo{volume}{D79}},
  \bibinfo{pages}{015014} (\bibinfo{year}{2009}), \eprint{0810.0713}.

\bibitem[{\citenamefont{Chen et~al.}(2008{\natexlab{a}})\citenamefont{Chen,
  Takahashi, and Yanagida}}]{Chen:2008yi}
\bibinfo{author}{\bibfnamefont{C.-R.} \bibnamefont{Chen}},
  \bibinfo{author}{\bibfnamefont{F.}~\bibnamefont{Takahashi}},
  \bibnamefont{and} \bibinfo{author}{\bibfnamefont{T.~T.}
  \bibnamefont{Yanagida}} (\bibinfo{year}{2008}{\natexlab{a}}),
  \eprint{0809.0792}.

\bibitem[{\citenamefont{Nelson and Spitzer}(2008)}]{Nelson:2008hj}
\bibinfo{author}{\bibfnamefont{A.~E.} \bibnamefont{Nelson}} \bibnamefont{and}
  \bibinfo{author}{\bibfnamefont{C.}~\bibnamefont{Spitzer}}
  (\bibinfo{year}{2008}), \eprint{0810.5167}.

\bibitem[{\citenamefont{Cholis et~al.}(2008{\natexlab{a}})\citenamefont{Cholis,
  Finkbeiner, Goodenough, and Weiner}}]{Cholis:2008qq}
\bibinfo{author}{\bibfnamefont{I.}~\bibnamefont{Cholis}},
  \bibinfo{author}{\bibfnamefont{D.~P.} \bibnamefont{Finkbeiner}},
  \bibinfo{author}{\bibfnamefont{L.}~\bibnamefont{Goodenough}},
  \bibnamefont{and} \bibinfo{author}{\bibfnamefont{N.}~\bibnamefont{Weiner}}
  (\bibinfo{year}{2008}{\natexlab{a}}), \eprint{0810.5344}.

\bibitem[{\citenamefont{Nomura and Thaler}(2008)}]{Nomura:2008ru}
\bibinfo{author}{\bibfnamefont{Y.}~\bibnamefont{Nomura}} \bibnamefont{and}
  \bibinfo{author}{\bibfnamefont{J.}~\bibnamefont{Thaler}}
  (\bibinfo{year}{2008}), \eprint{0810.5397}.

\bibitem[{\citenamefont{Harnik and Kribs}(2008)}]{Harnik:2008uu}
\bibinfo{author}{\bibfnamefont{R.}~\bibnamefont{Harnik}} \bibnamefont{and}
  \bibinfo{author}{\bibfnamefont{G.~D.} \bibnamefont{Kribs}}
  (\bibinfo{year}{2008}), \eprint{0810.5557}.

\bibitem[{\citenamefont{Bai and Han}(2009)}]{Bai:2008jt}
\bibinfo{author}{\bibfnamefont{Y.}~\bibnamefont{Bai}} \bibnamefont{and}
  \bibinfo{author}{\bibfnamefont{Z.}~\bibnamefont{Han}},
  \bibinfo{journal}{Phys. Rev.} \textbf{\bibinfo{volume}{D79}},
  \bibinfo{pages}{095023} (\bibinfo{year}{2009}), \eprint{0811.0387}.

\bibitem[{\citenamefont{Fox and Poppitz}(2009)}]{Fox:2008kb}
\bibinfo{author}{\bibfnamefont{P.~J.} \bibnamefont{Fox}} \bibnamefont{and}
  \bibinfo{author}{\bibfnamefont{E.}~\bibnamefont{Poppitz}},
  \bibinfo{journal}{Phys. Rev.} \textbf{\bibinfo{volume}{D79}},
  \bibinfo{pages}{083528} (\bibinfo{year}{2009}), \eprint{0811.0399}.

\bibitem[{\citenamefont{Ponton and Randall}(2009)}]{Ponton:2008zv}
\bibinfo{author}{\bibfnamefont{E.}~\bibnamefont{Ponton}} \bibnamefont{and}
  \bibinfo{author}{\bibfnamefont{L.}~\bibnamefont{Randall}},
  \bibinfo{journal}{JHEP} \textbf{\bibinfo{volume}{04}}, \bibinfo{pages}{080}
  (\bibinfo{year}{2009}), \eprint{0811.1029}.

\bibitem[{\citenamefont{Chen et~al.}(2008{\natexlab{b}})\citenamefont{Chen,
  Takahashi, and Yanagida}}]{Chen:2008qs}
\bibinfo{author}{\bibfnamefont{C.-R.} \bibnamefont{Chen}},
  \bibinfo{author}{\bibfnamefont{F.}~\bibnamefont{Takahashi}},
  \bibnamefont{and} \bibinfo{author}{\bibfnamefont{T.~T.}
  \bibnamefont{Yanagida}} (\bibinfo{year}{2008}{\natexlab{b}}),
  \eprint{0811.3357}.

\bibitem[{\citenamefont{Ibe et~al.}(2009)\citenamefont{Ibe, Murayama, and
  Yanagida}}]{Ibe:2008ye}
\bibinfo{author}{\bibfnamefont{M.}~\bibnamefont{Ibe}},
  \bibinfo{author}{\bibfnamefont{H.}~\bibnamefont{Murayama}}, \bibnamefont{and}
  \bibinfo{author}{\bibfnamefont{T.~T.} \bibnamefont{Yanagida}},
  \bibinfo{journal}{Phys. Rev.} \textbf{\bibinfo{volume}{D79}},
  \bibinfo{pages}{095009} (\bibinfo{year}{2009}), \eprint{0812.0072}.

\bibitem[{\citenamefont{Chun and Park}(2009)}]{Chun:2008by}
\bibinfo{author}{\bibfnamefont{E.~J.} \bibnamefont{Chun}} \bibnamefont{and}
  \bibinfo{author}{\bibfnamefont{J.-C.} \bibnamefont{Park}},
  \bibinfo{journal}{JCAP} \textbf{\bibinfo{volume}{0902}}, \bibinfo{pages}{026}
  (\bibinfo{year}{2009}), \eprint{0812.0308}.

\bibitem[{\citenamefont{Arvanitaki et~al.}(2008)}]{Arvanitaki:2008hq}
\bibinfo{author}{\bibfnamefont{A.}~\bibnamefont{Arvanitaki}}
  \bibnamefont{et~al.} (\bibinfo{year}{2008}), \eprint{0812.2075}.

\bibitem[{\citenamefont{Grajek et~al.}(2008)\citenamefont{Grajek, Kane, Phalen,
  Pierce, and Watson}}]{Grajek:2008pg}
\bibinfo{author}{\bibfnamefont{P.}~\bibnamefont{Grajek}},
  \bibinfo{author}{\bibfnamefont{G.}~\bibnamefont{Kane}},
  \bibinfo{author}{\bibfnamefont{D.}~\bibnamefont{Phalen}},
  \bibinfo{author}{\bibfnamefont{A.}~\bibnamefont{Pierce}}, \bibnamefont{and}
  \bibinfo{author}{\bibfnamefont{S.}~\bibnamefont{Watson}}
  (\bibinfo{year}{2008}), \eprint{0812.4555}.

\bibitem[{\citenamefont{Shirai et~al.}(2009)\citenamefont{Shirai, Takahashi,
  and Yanagida}}]{Shirai:2009kh}
\bibinfo{author}{\bibfnamefont{S.}~\bibnamefont{Shirai}},
  \bibinfo{author}{\bibfnamefont{F.}~\bibnamefont{Takahashi}},
  \bibnamefont{and} \bibinfo{author}{\bibfnamefont{T.~T.}
  \bibnamefont{Yanagida}} (\bibinfo{year}{2009}), \eprint{0902.4770}.

\bibitem[{\citenamefont{Mardon et~al.}(2009{\natexlab{a}})\citenamefont{Mardon,
  Nomura, and Thaler}}]{Mardon:2009gw}
\bibinfo{author}{\bibfnamefont{J.}~\bibnamefont{Mardon}},
  \bibinfo{author}{\bibfnamefont{Y.}~\bibnamefont{Nomura}}, \bibnamefont{and}
  \bibinfo{author}{\bibfnamefont{J.}~\bibnamefont{Thaler}}
  (\bibinfo{year}{2009}{\natexlab{a}}), \eprint{0905.3749}.

\bibitem[{\citenamefont{Serpico and Zaharijas}(2008)}]{Serpico:2008ga}
\bibinfo{author}{\bibfnamefont{P.~D.} \bibnamefont{Serpico}} \bibnamefont{and}
  \bibinfo{author}{\bibfnamefont{G.}~\bibnamefont{Zaharijas}},
  \bibinfo{journal}{Astropart. Phys.} \textbf{\bibinfo{volume}{29}},
  \bibinfo{pages}{380} (\bibinfo{year}{2008}), \eprint{0802.3245}.

\bibitem[{\citenamefont{Martinez et~al.}(2009)\citenamefont{Martinez, Bullock,
  Kaplinghat, Strigari, and Trotta}}]{Martinez:2009jh}
\bibinfo{author}{\bibfnamefont{G.~D.} \bibnamefont{Martinez}},
  \bibinfo{author}{\bibfnamefont{J.~S.} \bibnamefont{Bullock}},
  \bibinfo{author}{\bibfnamefont{M.}~\bibnamefont{Kaplinghat}},
  \bibinfo{author}{\bibfnamefont{L.~E.} \bibnamefont{Strigari}},
  \bibnamefont{and} \bibinfo{author}{\bibfnamefont{R.}~\bibnamefont{Trotta}},
  \bibinfo{journal}{JCAP} \textbf{\bibinfo{volume}{0906}}, \bibinfo{pages}{014}
  (\bibinfo{year}{2009}), \eprint{0902.4715}.

\bibitem[{\citenamefont{Baltz et~al.}(2000)\citenamefont{Baltz, Briot, Salati,
  Taillet, and Silk}}]{Baltz:1999ra}
\bibinfo{author}{\bibfnamefont{E.~A.} \bibnamefont{Baltz}},
  \bibinfo{author}{\bibfnamefont{C.}~\bibnamefont{Briot}},
  \bibinfo{author}{\bibfnamefont{P.}~\bibnamefont{Salati}},
  \bibinfo{author}{\bibfnamefont{R.}~\bibnamefont{Taillet}}, \bibnamefont{and}
  \bibinfo{author}{\bibfnamefont{J.}~\bibnamefont{Silk}},
  \bibinfo{journal}{Phys. Rev.} \textbf{\bibinfo{volume}{D61}},
  \bibinfo{pages}{023514} (\bibinfo{year}{2000}), \eprint{astro-ph/9909112}.

\bibitem[{\citenamefont{Kuhlen et~al.}(2008)\citenamefont{Kuhlen, Diemand, and
  Madau}}]{Kuhlen:2008aw}
\bibinfo{author}{\bibfnamefont{M.}~\bibnamefont{Kuhlen}},
  \bibinfo{author}{\bibfnamefont{J.}~\bibnamefont{Diemand}}, \bibnamefont{and}
  \bibinfo{author}{\bibfnamefont{P.}~\bibnamefont{Madau}}
  (\bibinfo{year}{2008}), \eprint{0805.4416}.

\bibitem[{\citenamefont{Baltz et~al.}(2008)}]{Baltz:2008wd}
\bibinfo{author}{\bibfnamefont{E.~A.} \bibnamefont{Baltz}}
  \bibnamefont{et~al.}, \bibinfo{journal}{JCAP}
  \textbf{\bibinfo{volume}{0807}}, \bibinfo{pages}{013} (\bibinfo{year}{2008}),
  \eprint{0806.2911}.

\bibitem[{\citenamefont{Bergstrom et~al.}(2001)\citenamefont{Bergstrom, Edsjo,
  and Ullio}}]{Bergstrom:2001jj}
\bibinfo{author}{\bibfnamefont{L.}~\bibnamefont{Bergstrom}},
  \bibinfo{author}{\bibfnamefont{J.}~\bibnamefont{Edsjo}}, \bibnamefont{and}
  \bibinfo{author}{\bibfnamefont{P.}~\bibnamefont{Ullio}},
  \bibinfo{journal}{Phys. Rev. Lett.} \textbf{\bibinfo{volume}{87}},
  \bibinfo{pages}{251301} (\bibinfo{year}{2001}), \eprint{astro-ph/0105048}.

\bibitem[{\citenamefont{Elsaesser and Mannheim}(2005)}]{Elsaesser:2004ap}
\bibinfo{author}{\bibfnamefont{D.}~\bibnamefont{Elsaesser}} \bibnamefont{and}
  \bibinfo{author}{\bibfnamefont{K.}~\bibnamefont{Mannheim}},
  \bibinfo{journal}{Phys. Rev. Lett.} \textbf{\bibinfo{volume}{94}},
  \bibinfo{pages}{171302} (\bibinfo{year}{2005}), \eprint{astro-ph/0405235}.

\bibitem[{\citenamefont{Kuhlen}(2009)}]{Kuhlen:2009jv}
\bibinfo{author}{\bibfnamefont{M.}~\bibnamefont{Kuhlen}}
  (\bibinfo{year}{2009}), \eprint{0906.1822}.

\bibitem[{\citenamefont{Siegal-Gaskins}(2008)}]{SiegalGaskins:2008ge}
\bibinfo{author}{\bibfnamefont{J.~M.} \bibnamefont{Siegal-Gaskins}},
  \bibinfo{journal}{JCAP} \textbf{\bibinfo{volume}{0810}}, \bibinfo{pages}{040}
  (\bibinfo{year}{2008}), \eprint{0807.1328}.

\bibitem[{\citenamefont{Dodelson et~al.}(2008)\citenamefont{Dodelson, Hooper,
  and Serpico}}]{Dodelson:2007gd}
\bibinfo{author}{\bibfnamefont{S.}~\bibnamefont{Dodelson}},
  \bibinfo{author}{\bibfnamefont{D.}~\bibnamefont{Hooper}}, \bibnamefont{and}
  \bibinfo{author}{\bibfnamefont{P.~D.} \bibnamefont{Serpico}},
  \bibinfo{journal}{Phys. Rev.} \textbf{\bibinfo{volume}{D77}},
  \bibinfo{pages}{063512} (\bibinfo{year}{2008}), \eprint{0711.4621}.

\bibitem[{\citenamefont{Pieri et~al.}(2009)\citenamefont{Pieri, Lavalle,
  Bertone, and Branchini}}]{Pieri:2009je}
\bibinfo{author}{\bibfnamefont{L.}~\bibnamefont{Pieri}},
  \bibinfo{author}{\bibfnamefont{J.}~\bibnamefont{Lavalle}},
  \bibinfo{author}{\bibfnamefont{G.}~\bibnamefont{Bertone}}, \bibnamefont{and}
  \bibinfo{author}{\bibfnamefont{E.}~\bibnamefont{Branchini}}
  (\bibinfo{year}{2009}), \eprint{0908.0195}.

\bibitem[{\citenamefont{Springel et~al.}(2008)}]{Springel:2008by}
\bibinfo{author}{\bibfnamefont{V.}~\bibnamefont{Springel}} \bibnamefont{et~al.}
  (\bibinfo{year}{2008}), \eprint{0809.0894}.

\bibitem[{\citenamefont{Cholis et~al.}(2009)}]{Cholis:2009gv}
\bibinfo{author}{\bibfnamefont{I.}~\bibnamefont{Cholis}} \bibnamefont{et~al.}
  (\bibinfo{year}{2009}), \eprint{0907.3953}.

\bibitem[{\citenamefont{Abdo et~al.}(2010)}]{Abdo:2010dk}
\bibinfo{author}{\bibfnamefont{A.~A.} \bibnamefont{Abdo}} \bibnamefont{et~al.}
  (\bibinfo{collaboration}{Fermi-LAT}), \bibinfo{journal}{JCAP}
  \textbf{\bibinfo{volume}{1004}}, \bibinfo{pages}{014} (\bibinfo{year}{2010}),
  \eprint{1002.4415}.

\bibitem[{\citenamefont{Cholis et~al.}(2008{\natexlab{b}})\citenamefont{Cholis,
  Dobler, Finkbeiner, Goodenough, and Weiner}}]{Cholis:2008wq}
\bibinfo{author}{\bibfnamefont{I.}~\bibnamefont{Cholis}},
  \bibinfo{author}{\bibfnamefont{G.}~\bibnamefont{Dobler}},
  \bibinfo{author}{\bibfnamefont{D.~P.} \bibnamefont{Finkbeiner}},
  \bibinfo{author}{\bibfnamefont{L.}~\bibnamefont{Goodenough}},
  \bibnamefont{and} \bibinfo{author}{\bibfnamefont{N.}~\bibnamefont{Weiner}}
  (\bibinfo{year}{2008}{\natexlab{b}}), \eprint{0811.3641}.

\bibitem[{\citenamefont{Zhang et~al.}(2009)}]{Zhang:2008tb}
\bibinfo{author}{\bibfnamefont{J.}~\bibnamefont{Zhang}} \bibnamefont{et~al.},
  \bibinfo{journal}{Phys. Rev.} \textbf{\bibinfo{volume}{D80}},
  \bibinfo{pages}{023007} (\bibinfo{year}{2009}), \eprint{0812.0522}.

\bibitem[{\citenamefont{Borriello et~al.}(2009)\citenamefont{Borriello, Cuoco,
  and Miele}}]{Borriello:2009fa}
\bibinfo{author}{\bibfnamefont{E.}~\bibnamefont{Borriello}},
  \bibinfo{author}{\bibfnamefont{A.}~\bibnamefont{Cuoco}}, \bibnamefont{and}
  \bibinfo{author}{\bibfnamefont{G.}~\bibnamefont{Miele}},
  \bibinfo{journal}{Astrophys. J.} \textbf{\bibinfo{volume}{699}},
  \bibinfo{pages}{L59} (\bibinfo{year}{2009}), \eprint{0903.1852}.

\bibitem[{\citenamefont{Cirelli and Panci}(2009)}]{Cirelli:2009vg}
\bibinfo{author}{\bibfnamefont{M.}~\bibnamefont{Cirelli}} \bibnamefont{and}
  \bibinfo{author}{\bibfnamefont{P.}~\bibnamefont{Panci}}
  (\bibinfo{year}{2009}), \eprint{0904.3830}.

\bibitem[{\citenamefont{Regis and Ullio}(2009)}]{Regis:2009md}
\bibinfo{author}{\bibfnamefont{M.}~\bibnamefont{Regis}} \bibnamefont{and}
  \bibinfo{author}{\bibfnamefont{P.}~\bibnamefont{Ullio}}
  (\bibinfo{year}{2009}), \eprint{0904.4645}.

\bibitem[{\citenamefont{Belikov and Hooper}(2009)}]{Belikov:2009cx}
\bibinfo{author}{\bibfnamefont{A.~V.} \bibnamefont{Belikov}} \bibnamefont{and}
  \bibinfo{author}{\bibfnamefont{D.}~\bibnamefont{Hooper}}
  (\bibinfo{year}{2009}), \eprint{0906.2251}.

\bibitem[{\citenamefont{Meade et~al.}(2009{\natexlab{a}})\citenamefont{Meade,
  Papucci, Strumia, and Volansky}}]{Meade:2009iu}
\bibinfo{author}{\bibfnamefont{P.}~\bibnamefont{Meade}},
  \bibinfo{author}{\bibfnamefont{M.}~\bibnamefont{Papucci}},
  \bibinfo{author}{\bibfnamefont{A.}~\bibnamefont{Strumia}}, \bibnamefont{and}
  \bibinfo{author}{\bibfnamefont{T.}~\bibnamefont{Volansky}}
  (\bibinfo{year}{2009}{\natexlab{a}}), \eprint{0905.0480}.

\bibitem[{\citenamefont{Beacom et~al.}(2005)\citenamefont{Beacom, Bell, and
  Bertone}}]{Beacom:2004pe}
\bibinfo{author}{\bibfnamefont{J.~F.} \bibnamefont{Beacom}},
  \bibinfo{author}{\bibfnamefont{N.~F.} \bibnamefont{Bell}}, \bibnamefont{and}
  \bibinfo{author}{\bibfnamefont{G.}~\bibnamefont{Bertone}},
  \bibinfo{journal}{Phys. Rev. Lett.} \textbf{\bibinfo{volume}{94}},
  \bibinfo{pages}{171301} (\bibinfo{year}{2005}), \eprint{astro-ph/0409403}.

\bibitem[{\citenamefont{Bergstrom et~al.}(2005)\citenamefont{Bergstrom,
  Bringmann, Eriksson, and Gustafsson}}]{Bergstrom:2004cy}
\bibinfo{author}{\bibfnamefont{L.}~\bibnamefont{Bergstrom}},
  \bibinfo{author}{\bibfnamefont{T.}~\bibnamefont{Bringmann}},
  \bibinfo{author}{\bibfnamefont{M.}~\bibnamefont{Eriksson}}, \bibnamefont{and}
  \bibinfo{author}{\bibfnamefont{M.}~\bibnamefont{Gustafsson}},
  \bibinfo{journal}{Phys. Rev. Lett.} \textbf{\bibinfo{volume}{94}},
  \bibinfo{pages}{131301} (\bibinfo{year}{2005}), \eprint{astro-ph/0410359}.

\bibitem[{\citenamefont{Birkedal et~al.}(2005)\citenamefont{Birkedal, Matchev,
  Perelstein, and Spray}}]{Birkedal:2005ep}
\bibinfo{author}{\bibfnamefont{A.}~\bibnamefont{Birkedal}},
  \bibinfo{author}{\bibfnamefont{K.~T.} \bibnamefont{Matchev}},
  \bibinfo{author}{\bibfnamefont{M.}~\bibnamefont{Perelstein}},
  \bibnamefont{and} \bibinfo{author}{\bibfnamefont{A.}~\bibnamefont{Spray}}
  (\bibinfo{year}{2005}), \eprint{hep-ph/0507194}.

\bibitem[{\citenamefont{Mack et~al.}(2008)\citenamefont{Mack, Jacques, Beacom,
  Bell, and Yuksel}}]{Mack:2008wu}
\bibinfo{author}{\bibfnamefont{G.~D.} \bibnamefont{Mack}},
  \bibinfo{author}{\bibfnamefont{T.~D.} \bibnamefont{Jacques}},
  \bibinfo{author}{\bibfnamefont{J.~F.} \bibnamefont{Beacom}},
  \bibinfo{author}{\bibfnamefont{N.~F.} \bibnamefont{Bell}}, \bibnamefont{and}
  \bibinfo{author}{\bibfnamefont{H.}~\bibnamefont{Yuksel}},
  \bibinfo{journal}{Phys. Rev.} \textbf{\bibinfo{volume}{D78}},
  \bibinfo{pages}{063542} (\bibinfo{year}{2008}), \eprint{0803.0157}.

\bibitem[{\citenamefont{Bergstrom et~al.}(2008)\citenamefont{Bergstrom,
  Bringmann, and Edsjo}}]{Bergstrom:2008gr}
\bibinfo{author}{\bibfnamefont{L.}~\bibnamefont{Bergstrom}},
  \bibinfo{author}{\bibfnamefont{T.}~\bibnamefont{Bringmann}},
  \bibnamefont{and} \bibinfo{author}{\bibfnamefont{J.}~\bibnamefont{Edsjo}},
  \bibinfo{journal}{Phys. Rev.} \textbf{\bibinfo{volume}{D78}},
  \bibinfo{pages}{103520} (\bibinfo{year}{2008}), \eprint{0808.3725}.

\bibitem[{\citenamefont{Bertone et~al.}(2009)\citenamefont{Bertone, Cirelli,
  Strumia, and Taoso}}]{Bertone:2008xr}
\bibinfo{author}{\bibfnamefont{G.}~\bibnamefont{Bertone}},
  \bibinfo{author}{\bibfnamefont{M.}~\bibnamefont{Cirelli}},
  \bibinfo{author}{\bibfnamefont{A.}~\bibnamefont{Strumia}}, \bibnamefont{and}
  \bibinfo{author}{\bibfnamefont{M.}~\bibnamefont{Taoso}},
  \bibinfo{journal}{JCAP} \textbf{\bibinfo{volume}{0903}}, \bibinfo{pages}{009}
  (\bibinfo{year}{2009}), \eprint{0811.3744}.

\bibitem[{\citenamefont{Bergstrom et~al.}(2009)\citenamefont{Bergstrom,
  Bertone, Bringmann, Edsjo, and Taoso}}]{Bergstrom:2008ag}
\bibinfo{author}{\bibfnamefont{L.}~\bibnamefont{Bergstrom}},
  \bibinfo{author}{\bibfnamefont{G.}~\bibnamefont{Bertone}},
  \bibinfo{author}{\bibfnamefont{T.}~\bibnamefont{Bringmann}},
  \bibinfo{author}{\bibfnamefont{J.}~\bibnamefont{Edsjo}}, \bibnamefont{and}
  \bibinfo{author}{\bibfnamefont{M.}~\bibnamefont{Taoso}},
  \bibinfo{journal}{Phys. Rev.} \textbf{\bibinfo{volume}{D79}},
  \bibinfo{pages}{081303} (\bibinfo{year}{2009}), \eprint{0812.3895}.

\bibitem[{\citenamefont{Meade et~al.}(2009{\natexlab{b}})\citenamefont{Meade,
  Papucci, and Volansky}}]{Meade:2009rb}
\bibinfo{author}{\bibfnamefont{P.}~\bibnamefont{Meade}},
  \bibinfo{author}{\bibfnamefont{M.}~\bibnamefont{Papucci}}, \bibnamefont{and}
  \bibinfo{author}{\bibfnamefont{T.}~\bibnamefont{Volansky}}
  (\bibinfo{year}{2009}{\natexlab{b}}), \eprint{0901.2925}.

\bibitem[{\citenamefont{Mardon et~al.}(2009{\natexlab{b}})\citenamefont{Mardon,
  Nomura, Stolarski, and Thaler}}]{Mardon:2009rc}
\bibinfo{author}{\bibfnamefont{J.}~\bibnamefont{Mardon}},
  \bibinfo{author}{\bibfnamefont{Y.}~\bibnamefont{Nomura}},
  \bibinfo{author}{\bibfnamefont{D.}~\bibnamefont{Stolarski}},
  \bibnamefont{and} \bibinfo{author}{\bibfnamefont{J.}~\bibnamefont{Thaler}}
  (\bibinfo{year}{2009}{\natexlab{b}}), \eprint{0901.2926}.

\bibitem[{\citenamefont{Strong et~al.}(2007)\citenamefont{Strong, Moskalenko,
  and Ptuskin}}]{Strong:2007nh}
\bibinfo{author}{\bibfnamefont{A.~W.} \bibnamefont{Strong}},
  \bibinfo{author}{\bibfnamefont{I.~V.} \bibnamefont{Moskalenko}},
  \bibnamefont{and} \bibinfo{author}{\bibfnamefont{V.~S.}
  \bibnamefont{Ptuskin}}, \bibinfo{journal}{Ann. Rev. Nucl. Part. Sci.}
  \textbf{\bibinfo{volume}{57}}, \bibinfo{pages}{285} (\bibinfo{year}{2007}),
  \eprint{astro-ph/0701517}.

\bibitem[{\citenamefont{Moskalenko and Strong}(1999)}]{Moskalenko:1999sb}
\bibinfo{author}{\bibfnamefont{I.~V.} \bibnamefont{Moskalenko}}
  \bibnamefont{and} \bibinfo{author}{\bibfnamefont{A.~W.}
  \bibnamefont{Strong}}, \bibinfo{journal}{Phys. Rev.}
  \textbf{\bibinfo{volume}{D60}}, \bibinfo{pages}{063003}
  (\bibinfo{year}{1999}), \eprint{astro-ph/9905283}.

\bibitem[{\citenamefont{Strong and Moskalenko}(1999)}]{Strong:1999sv}
\bibinfo{author}{\bibfnamefont{A.~W.} \bibnamefont{Strong}} \bibnamefont{and}
  \bibinfo{author}{\bibfnamefont{I.~V.} \bibnamefont{Moskalenko}}
  (\bibinfo{year}{1999}), \eprint{astro-ph/9906228}.

\bibitem[{\citenamefont{Strong et~al.}(2006)\citenamefont{Strong, Moskalenko,
  and Ptuskin}}]{Galprop1}
\bibinfo{author}{\bibfnamefont{A.~W.} \bibnamefont{Strong}},
  \bibinfo{author}{\bibfnamefont{I.~V.} \bibnamefont{Moskalenko}},
  \bibnamefont{and} \bibinfo{author}{\bibfnamefont{V.~S.}
  \bibnamefont{Ptuskin}}, \emph{\bibinfo{title}{GALPROP C++ v.50: Explanatory
  Supplement}} (\bibinfo{year}{2006}).

\bibitem[{\citenamefont{Porter and Strong}(2005)}]{Porter:2005qx}
\bibinfo{author}{\bibfnamefont{T.~A.} \bibnamefont{Porter}} \bibnamefont{and}
  \bibinfo{author}{\bibfnamefont{A.~W.} \bibnamefont{Strong}}
  (\bibinfo{year}{2005}), \eprint{astro-ph/0507119}.

\bibitem[{\citenamefont{Strong and Moskalenko}(1998)}]{Strong:1998pw}
\bibinfo{author}{\bibfnamefont{A.~W.} \bibnamefont{Strong}} \bibnamefont{and}
  \bibinfo{author}{\bibfnamefont{I.~V.} \bibnamefont{Moskalenko}},
  \bibinfo{journal}{Astrophys. J.} \textbf{\bibinfo{volume}{509}},
  \bibinfo{pages}{212} (\bibinfo{year}{1998}), \eprint{astro-ph/9807150}.

\bibitem[{\citenamefont{{Merritt} et~al.}(2005)\citenamefont{{Merritt},
  {Navarro}, {Ludlow}, and {Jenkins}}}]{Merritt:2005}
\bibinfo{author}{\bibfnamefont{D.}~\bibnamefont{{Merritt}}},
  \bibinfo{author}{\bibfnamefont{J.~F.} \bibnamefont{{Navarro}}},
  \bibinfo{author}{\bibfnamefont{A.}~\bibnamefont{{Ludlow}}}, \bibnamefont{and}
  \bibinfo{author}{\bibfnamefont{A.}~\bibnamefont{{Jenkins}}},
  \bibinfo{journal}{\apjl} \textbf{\bibinfo{volume}{624}}, \bibinfo{pages}{L85}
  (\bibinfo{year}{2005}), \eprint{astro-ph/0502515}.

\bibitem[{\citenamefont{Hisano et~al.}(2005)\citenamefont{Hisano, Matsumoto,
  Nojiri, and Saito}}]{Hisano:2004ds}
\bibinfo{author}{\bibfnamefont{J.}~\bibnamefont{Hisano}},
  \bibinfo{author}{\bibfnamefont{S.}~\bibnamefont{Matsumoto}},
  \bibinfo{author}{\bibfnamefont{M.~M.} \bibnamefont{Nojiri}},
  \bibnamefont{and} \bibinfo{author}{\bibfnamefont{O.}~\bibnamefont{Saito}},
  \bibinfo{journal}{Phys. Rev.} \textbf{\bibinfo{volume}{D71}},
  \bibinfo{pages}{063528} (\bibinfo{year}{2005}), \eprint{hep-ph/0412403}.

\bibitem[{\citenamefont{Cirelli et~al.}(2007)\citenamefont{Cirelli, Strumia,
  and Tamburini}}]{Cirelli:2007xd}
\bibinfo{author}{\bibfnamefont{M.}~\bibnamefont{Cirelli}},
  \bibinfo{author}{\bibfnamefont{A.}~\bibnamefont{Strumia}}, \bibnamefont{and}
  \bibinfo{author}{\bibfnamefont{M.}~\bibnamefont{Tamburini}},
  \bibinfo{journal}{Nucl. Phys.} \textbf{\bibinfo{volume}{B787}},
  \bibinfo{pages}{152} (\bibinfo{year}{2007}), \eprint{0706.4071}.

\bibitem[{\citenamefont{Arkani-Hamed and Weiner}(2008)}]{ArkaniHamed:2008qp}
\bibinfo{author}{\bibfnamefont{N.}~\bibnamefont{Arkani-Hamed}}
  \bibnamefont{and} \bibinfo{author}{\bibfnamefont{N.}~\bibnamefont{Weiner}},
  \bibinfo{journal}{JHEP} \textbf{\bibinfo{volume}{12}}, \bibinfo{pages}{104}
  (\bibinfo{year}{2008}), \eprint{0810.0714}.

\bibitem[{\citenamefont{Adriani et~al.}(2008{\natexlab{b}})}]{Adriani:2008zq}
\bibinfo{author}{\bibfnamefont{O.}~\bibnamefont{Adriani}} \bibnamefont{et~al.}
  (\bibinfo{year}{2008}{\natexlab{b}}), \eprint{0810.4994}.

\bibitem[{\citenamefont{Hunter et~al.}(1997)}]{Hunter:1997we}
\bibinfo{author}{\bibfnamefont{S.~D.} \bibnamefont{Hunter}}
  \bibnamefont{et~al.}, \bibinfo{journal}{Astrophys. J.}
  \textbf{\bibinfo{volume}{481}}, \bibinfo{pages}{205} (\bibinfo{year}{1997}).

\bibitem[{\citenamefont{Sreekumar et~al.}(1998)}]{Sreekumar:1997un}
\bibinfo{author}{\bibfnamefont{P.}~\bibnamefont{Sreekumar}}
  \bibnamefont{et~al.} (\bibinfo{collaboration}{EGRET}),
  \bibinfo{journal}{Astrophys. J.} \textbf{\bibinfo{volume}{494}},
  \bibinfo{pages}{523} (\bibinfo{year}{1998}), \eprint{astro-ph/9709257}.

\bibitem[{\citenamefont{Aharonian et~al.}(2004)}]{Aharonian:2004wa}
\bibinfo{author}{\bibfnamefont{F.}~\bibnamefont{Aharonian}}
  \bibnamefont{et~al.} (\bibinfo{collaboration}{The HESS}),
  \bibinfo{journal}{Astron. Astrophys.} \textbf{\bibinfo{volume}{425}},
  \bibinfo{pages}{L13} (\bibinfo{year}{2004}), \eprint{astro-ph/0408145}.

\bibitem[{\citenamefont{Aharonian et~al.}(2006)}]{Aharonian:2006wh}
\bibinfo{author}{\bibfnamefont{F.}~\bibnamefont{Aharonian}}
  \bibnamefont{et~al.} (\bibinfo{collaboration}{H.E.S.S.}),
  \bibinfo{journal}{Phys. Rev. Lett.} \textbf{\bibinfo{volume}{97}},
  \bibinfo{pages}{221102} (\bibinfo{year}{2006}), \eprint{astro-ph/0610509}.

\bibitem[{\citenamefont{Bell and Jacques}(2008)}]{Bell:2008vx}
\bibinfo{author}{\bibfnamefont{N.~F.} \bibnamefont{Bell}} \bibnamefont{and}
  \bibinfo{author}{\bibfnamefont{T.~D.} \bibnamefont{Jacques}}
  (\bibinfo{year}{2008}), \eprint{0811.0821}.

\bibitem[{\citenamefont{Ackermann}(2009)}]{ACKERMANNTALK}
\bibinfo{author}{\bibfnamefont{M.}~\bibnamefont{Ackermann}}
  (\bibinfo{collaboration}{Fermi-LAT}) (\bibinfo{year}{2009}), \eprint{Talk
  given at TeV Particle Astrophysics (TeVPA), July 13-17, 2009}.

\bibitem[{\citenamefont{Aguilar et~al.}(2007)}]{AMS}
\bibinfo{author}{\bibfnamefont{M.}~\bibnamefont{Aguilar}} \bibnamefont{et~al.}
  (\bibinfo{collaboration}{AMS-01}), \bibinfo{journal}{Phys. Lett.}
  \textbf{\bibinfo{volume}{B646}}, \bibinfo{pages}{145} (\bibinfo{year}{2007}),
  \eprint{astro-ph/0703154}.

\bibitem[{\citenamefont{Sanuki et~al.}(2000)}]{BESS}
\bibinfo{author}{\bibfnamefont{T.}~\bibnamefont{Sanuki}} \bibnamefont{et~al.},
  \bibinfo{journal}{Astrophys. J.} \textbf{\bibinfo{volume}{545}},
  \bibinfo{pages}{1135} (\bibinfo{year}{2000}), \eprint{astro-ph/0002481}.

\bibitem[{\citenamefont{{Menn} et~al.}(2000)\citenamefont{{Menn}, {Hof},
  {Reimer}, {Simon}, {Davis}, {Labrador}, {Mewaldt}, {Schindler}, {Barbier},
  {Christian} et~al.}}]{IMAX}
\bibinfo{author}{\bibfnamefont{W.}~\bibnamefont{{Menn}}},
  \bibinfo{author}{\bibfnamefont{M.}~\bibnamefont{{Hof}}},
  \bibinfo{author}{\bibfnamefont{O.}~\bibnamefont{{Reimer}}},
  \bibinfo{author}{\bibfnamefont{M.}~\bibnamefont{{Simon}}},
  \bibinfo{author}{\bibfnamefont{A.~J.} \bibnamefont{{Davis}}},
  \bibinfo{author}{\bibfnamefont{A.~W.} \bibnamefont{{Labrador}}},
  \bibinfo{author}{\bibfnamefont{R.~A.} \bibnamefont{{Mewaldt}}},
  \bibinfo{author}{\bibfnamefont{S.~M.} \bibnamefont{{Schindler}}},
  \bibinfo{author}{\bibfnamefont{L.~M.} \bibnamefont{{Barbier}}},
  \bibinfo{author}{\bibfnamefont{E.~R.} \bibnamefont{{Christian}}},
  \bibnamefont{et~al.}, \bibinfo{journal}{\apj} \textbf{\bibinfo{volume}{533}},
  \bibinfo{pages}{281} (\bibinfo{year}{2000}).

\bibitem[{\citenamefont{Barger et~al.}(2009)\citenamefont{Barger, Gao, Keung,
  Marfatia, and Shaughnessy}}]{Barger:2009yt}
\bibinfo{author}{\bibfnamefont{V.}~\bibnamefont{Barger}},
  \bibinfo{author}{\bibfnamefont{Y.}~\bibnamefont{Gao}},
  \bibinfo{author}{\bibfnamefont{W.~Y.} \bibnamefont{Keung}},
  \bibinfo{author}{\bibfnamefont{D.}~\bibnamefont{Marfatia}}, \bibnamefont{and}
  \bibinfo{author}{\bibfnamefont{G.}~\bibnamefont{Shaughnessy}},
  \bibinfo{journal}{Phys. Lett.} \textbf{\bibinfo{volume}{B678}},
  \bibinfo{pages}{283} (\bibinfo{year}{2009}), \eprint{0904.2001}.

\bibitem[{\citenamefont{Atwood et~al.}(2009)}]{Atwood:2009ez}
\bibinfo{author}{\bibfnamefont{W.~B.} \bibnamefont{Atwood}}
  \bibnamefont{et~al.} (\bibinfo{collaboration}{LAT}),
  \bibinfo{journal}{Astrophys. J.} \textbf{\bibinfo{volume}{697}},
  \bibinfo{pages}{1071} (\bibinfo{year}{2009}), \eprint{0902.1089}.

\bibitem[{\citenamefont{Simet and Hooper}(2009)}]{Simet:2009ne}
\bibinfo{author}{\bibfnamefont{M.}~\bibnamefont{Simet}} \bibnamefont{and}
  \bibinfo{author}{\bibfnamefont{D.}~\bibnamefont{Hooper}}
  (\bibinfo{year}{2009}), \eprint{0904.2398}.

\end{thebibliography}
\bibliographystyle{apsrev}

\end{document}